\documentclass[apj]{emulateapj}
\usepackage{graphicx}

\shorttitle{ALMA Observations of Henize 2-10}
\shortauthors{Johnson et al.}

\begin{document}

\submitted{Accepted ApJ, in press}
\title{The Association of Molecular Gas and Natal Super Star Clusters
  in Henize 2-10}

\author{Kelsey E. Johnson \altaffilmark{1,2}, Crystal L. Brogan
  \altaffilmark{3}, Remy Indebetouw \altaffilmark{1,3}, Leonardo Testi
  \altaffilmark{4,5}, David J. Wilner \altaffilmark{6}, Amy E. Reines
  \altaffilmark{7}, C.-H. Rosie Chen \altaffilmark{8}, Leonardo Vanzi \altaffilmark{9} }
\altaffiltext{1}{Department of Astronomy, University of Virginia,
    Charlottesville, VA 22904-4325, USA}

\altaffiltext{2}{Adjunct Astronomer, National Radio Astronomy Observatory, 520 Edgemont Road,
  Charlottesville, VA 22903, USA}

\altaffiltext{3}{National Radio Astronomy Observatory, 520 Edgemont Road,
  Charlottesville, VA 22903, USA}

\altaffiltext{4}{European Southern Observatory, D-85748 Garching bei M\"{u}nchen, Germany}

\altaffiltext{5}{INAF/Osservatorio Astrofisico di Arcetri, Largo E. Fermi 5, I-50125 Firenze, Italy}

\altaffiltext{6}{Harvard-Smithsonian Center for Astrophysics, 60 Garden St., Cambridge, MA 02138, USA}

\altaffiltext{7}{Department of Physics, Montana State University, Bozeman, MT 59717, USA}

\altaffiltext{8}{Max Planck Institute for Radio Astronomy, D-53121 Bonn, Germany}

\altaffiltext{9}{Department of Electrical Engineering and Center of Astro Engineering, Pontificia Universidad Catolica de Chile, Av. Vicuña Mackenna 4860 Santiago, Chile}

\begin{abstract}
  We present ALMA observations of the dwarf starburst galaxy He~2-10
  in combination with previous SMA CO observations to probe the
  molecular environments of natal super star clusters.  These
  observations include the HCO$^+$(1-0), HCN(1-0), HNC(1-0), and
  CCH(1-0) molecular lines, as well as 88~GHz continuum with a spatial
  resolution of $1''.7\times 1''.6$.  After correcting for the
  contribution from free-free emission to the 88~GHz continuum flux
  density ($\sim$ 60\% of the 88~GHz emission), we derive a total gas
  mass for He~2-10 of $M_{gas} = 4-6\times10^8$~M$_{\odot}$, roughly
  5-20\% of the dynamical mass.  Based on a principle component
  analysis, HCO$^+$ is found to be the best ``general'' tracer of
  molecular emission.  The line widths and luminosities of the CO
  emission suggests that the molecular clouds could either be as small
  as $\sim 8$~pc, or alternately have enhanced line widths.  The CO
  emission and 88~GHz continuum are anti-correlated, suggesting that
  either the dust and molecular gas are not cospatial, which could
  reflect the 88~GHz continuum is dominated by free-free emission.
  The CO and CCH emission are also relatively anti-correlated, which
  is consistent with the CCH being photo-enhanced, and/or the CO being
  dissociated in the regions near the natal super star clusters.  The
  molecular line ratios of regions containing the natal star clusters
  are different from the line ratios observed for regions elsewhere in
  the galaxy.  In particular, the regions with thermal radio emission
  all have CO(2-1)/HCO$^+(1-0) < 16$, and the HCO$^+$/CO ratio appears
  to be correlated with the evolutionary stage of the clusters.

\end{abstract}


\section{Introduction}
 \label{intro}

Millimeter observations of starburst galaxies have demonstrated
  that their spectra contain numerous molecular lines.  Over
  the last two decades spatially resolved observations using
  interferometers of relatively nearby systems $\lesssim 4$~Mpc have
  become prevalent (e.g. NGC~253, \citealt{garcia-burillo00} and
  IC~342, \citealt{meier05}).  Now with the addition of the Atacama
  Large Millimeter/submillimeter Array (ALMA) to our toolbox, detailed
  studies of molecular lines in even more distant galaxies are
  becoming commonplace.  The detected molecular lines can provide new
  insight into the physical conditions of their host galaxies,
  including temperatures, densities, and radiation fields.  Our goal
  in this program is to utilize the sensitivity and resolution of ALMA
  to investigate the physical conditions associated with the so-called
  ``super star clusters''.

Super star clusters (SSCs) have stellar densities exceeding $10^4$
stars pc$^{-3}$ in their cores and masses often exceeding
$10^5$~M$_\odot$.  As ``extreme'' objects, one might be tempted to
simply think of SSCs as ``cosmic curiosities''.  However, these
objects are not only fascinating because of their extreme nature, but
also because they hold clues to an important mode of star formation
during the time of galaxy assembly.  Specifically, most present-day
research strongly supports the idea that SSCs are the adolescent
precursors to the ancient globular clusters ubiquitous around massive
galaxies in the local universe today \citep[e.g.][and references
  therein]{harris03}.  Given their estimated infant mortality rates
\citep[possibly as high as 99\%;][]{fall01}, the production of SSCs
must have been truly prodigious in the early universe, making this
extreme type of star formation a critical (and perhaps dominant) mode
in the early evolution of today's massive galaxies.

SSCs are also important components of galaxies in the present-day
universe.  These clusters can have a major impact on their surrounding
interstellar medium (ISM), and in some cases even the intergalactic
medium (IGM).  Each of these clusters can host thousands of massive
stars with powerful stellar winds that will all die in a violent way
within a span of a few million years.  The collective effect of these
massive stars can cause tremendous outflows, ionize large volumes of
the ISM and IGM, and enrich vast amounts of interstellar material
\citep[e.g.][]{johnson00,heckman01,martin02}

Radio and infrared observations have identified {\it nascent} SSCs
\citep[e.g.][]{kj99,turner00}, still embedded in their birth material.
However, the physical conditions in these regions have been virtually
impossible to assess without sufficient millimeter capabilities.

\subsection{The Target Galaxy: Henize 2-10}
\begin{figure*}[t!]
\centering
\plotone{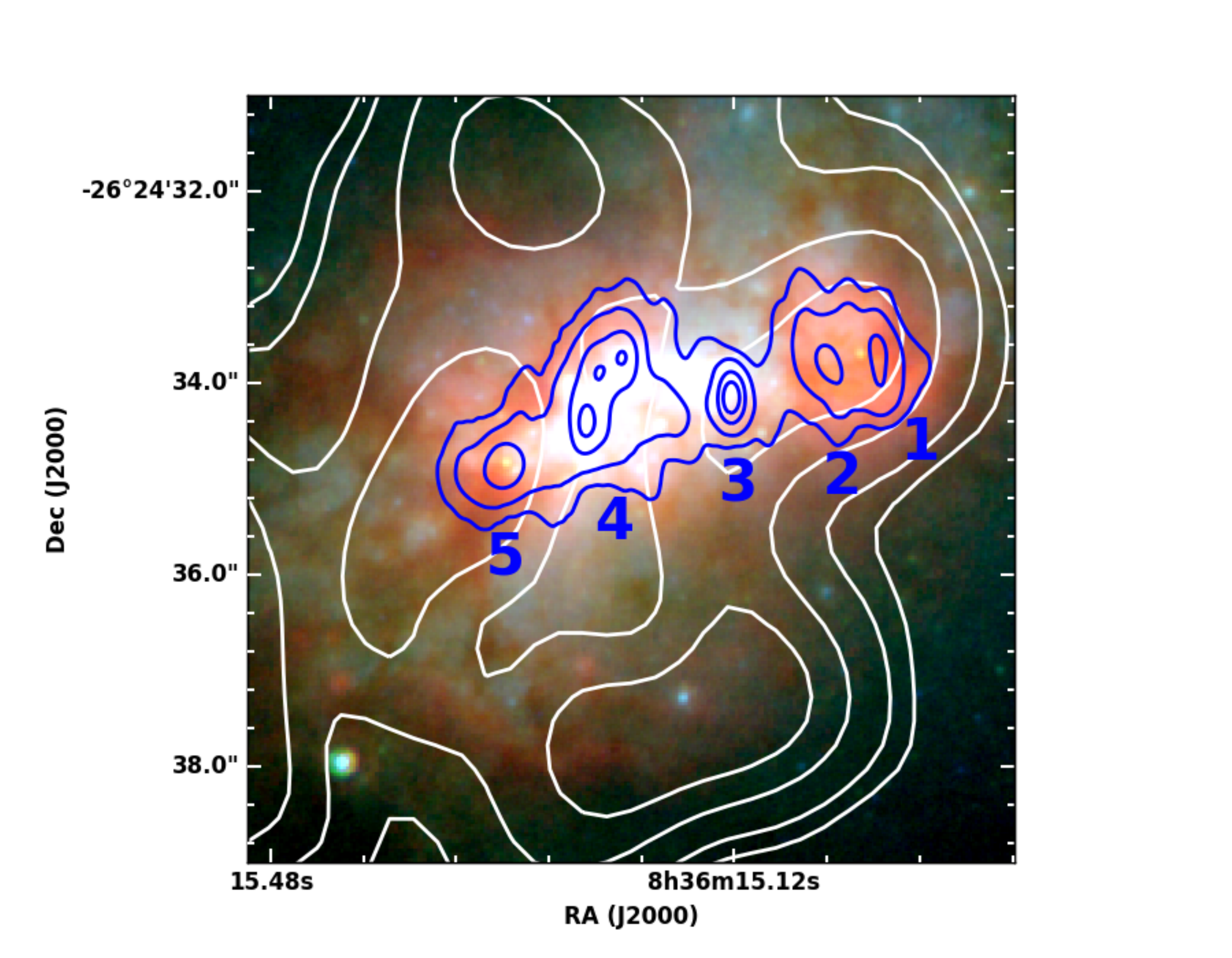}
\caption{The central region of He~2-10.  The underlying
  color image is from HST: red=F187N ($\sim$Pa$\alpha$), green=F814W
  ($\sim$I-band), and blue=F439W ($\sim$B-band).  CO(2-1) contours are
 shown in white and 3.6~cm contours are shown in blue.   The radio
 sources are labeled according to \citet{kj99}.  Source 3 is known to
 be an AGN \citep{reines11}. At the adopted distance of He~2-10 of 9~Mpc, 1$'' \approx 44$~pc. \label{plot_radio_cont}}
\end{figure*}

Henize 2-10 (He~2-10) is a remarkable blue compact dwarf galaxy that
has attracted astronomers' interest for decades, largely due to its high star
formation rate \citep[$\sim$0.7~M$_\odot$~yr$^{-1}$,][]{lee09} and
powerful outflows \citep[$> 360$~km~s$^{-1}$][]{johnson00}. He~2-10 is
the nearest ($D\sim 9$~Mpc) galaxy known to host {\it multiple} nascent
super star clusters detected in radio and infrared observations
\citep[see Figure~\ref{plot_radio_cont},][]{kj99,jk03,vacca02, cabanac05}. More
recently, a low-luminosity AGN was identified in this system, making
He~2-10 the first known example of a starbursting dwarf hosting a
massive black hole \citep{reines11, reines16}.

With its intense burst of star formation, irregular morphology (see
Figure~\ref{4panel}), large population of adolescent super star
clusters \citep{johnson00}, and low-luminosity AGN \citep{reines11},
many of the properties of He~2-10 are similar to those expected of
protogalaxies at high redshift during the early stages of hierarchical
galaxy assembly and globular cluster formation.  Especially important
for this study is the presence of multiple nascent super star clusters
detected through their thermal radio emission.  Given these
characteristics, He~2-10 is (arguably) one of the best available
nearby laboratories in which to study the extreme physical conditions
that lead to the formation of SSCs (and may have led to the formation
of globular clusters).  In this context it is also important to note
that, unlike most dwarf galaxies, He~2-10 actually has relatively high
metallicity of $12+$log(O/H)$=8.9$ inferred by \citet{kobulnicky99},
however the effect of non-primordial metallicities on star formation
remains poorly constrained.  Long wavelength observations are
essential to probe the birth environments of these clusters, and
(sub)millimeter observations also provide molecular diagnostics that
can be used to infer physical properties.

He~2-10 has a range of existing millimeter observations
\citep{santangelo09, vanzi09, imanishi07, meier01, kobulnicky95}.
The CO(1-0) observations of \citet{kobulnicky95} with a beam size of
$6''$ indicated a dynamically perturbed system, with the peak CO
emission offset by $\sim 100$~pc from the optical peak, and an
elongated feature roughly $30''$ ($=1.3$~kpc) in extent, which those
authors interpreted as a tidal feature, but is also consistent with an
infalling cloud.  Using CO(3-2), CO(2-1), and CO(1-0) observations
with a $\sim 22''$ beam, \citet{meier01} find a best-fit solution of
$n_{H2}>10^{3.5}$~cm$^{-3}$ and T$_K \sim 5-10$~K.  These results are
in apparent tension with those of \citet{bayet04}, who found T$_K
\approx 50-100$~K and $n_{H2}>10^{4}$~cm$^{-3}$ using higher level CO
transitions (up to CO(7-6)) with beam sizes ranging from $\sim 9
-30''$ (although these authors note that He~2-10 is point-like for all
lines).  \citet{imanishi07} observed He~2-10 with a beam of $10.''5
\times 5.''5$, and find the HCN(1-0)/HCO$^+$(1-0)$<0.6$, which is
relatively low compared to the AGN-dominated galaxies in their sample
\footnote{The massive black hole in He~2-10 is radiating significantly
  below its Eddington limit \citep[$\sim 10^{-6}$~L$_{\rm
    Edd}$][]{reines16}, and therefore we do not necessarily expect the
  line ratios to be similar to more luminous AGNs.}. \citet{vanzi09}
obtained APEX observations of He~2-10 in CO(3-2) and HCN(4-3) with a
$\sim 20''$ beam; they do not detect the CO elongation noted by
\citet{kobulnicky95}, but they clearly detect two velocity components
including a feature to the north-east.  \citet{santangelo09} obtained
significantly higher spatial resolution observations of He~2-10 than
previous studies, with a $1.''9 \times 1.''3$ beam from the SMA.
These observations revealed complex spatial structure, including a
bright CO feature toward the south-east, which could be associated
with the ``tail'' identified by \citet{kobulnicky95}.  In addition,
they find possible evidence for the association of compact molecular
clouds and super star clusters.

Figure~\ref{plot_radio_cont} shows the central region of He~2-10 with
the CO(2-1) and 3.6~cm radio emission overlaid in contours, and the radio
sources 1-5 \citep[as originally numbered by][]{kj99}.   Sources 1, 2,
3, and 5 are consistent with locations of CO emission.  Source 4
appears to fall in a CO valley, which is consistent with the slightly
non-thermal radio spectral index, and the inference of
\citet{cabanac05} that this region includes supernova remnants,
suggesting the presence of star clusters at least a few Myr
in age.     \citet{cabanac05} also determine that Sources 1, 2, and 5
likely have optical extinctions A$_V\gtrsim 10$, consistent with these
sources still being highly embedded in their birth material.
Source~3 has a non-thermal spectral index \citep{jk03}, and was
confirmed by \citet{reines11} to host a low-luminosity AGN.

In order to better understand the complex molecular emission in this
dwarf galaxy, we present sensitive new ALMA observations of the
HCO$^+$(1-0), HCN(1-0), HNC(1-0), and CCH(1-0) transitions together
with analysis of existing SMA CO(2-1) data.  

The paper is organized as follows: In \S\ref{obs} we give an
overview of the observations used in this analysis.
\S\ref{results}, we present the properties of the line and
continuum emission, including morphologies, strengths, line widths,
ratios, as well as a principle component analysis, a comparison to
single dish observations, and the inferred gas and dust mass.  In
\S\ref{discussion} we discuss the extent to which these
molecular line ratios may be tracking the evolution of super star
clusters, and further compare the results to those of other galaxies.
Finally, in \S\ref{summary}, we summarize the main points of this
paper.

\section{Observations}
 \label{obs}

\begin{figure*}
\centering
\includegraphics[width=6in]{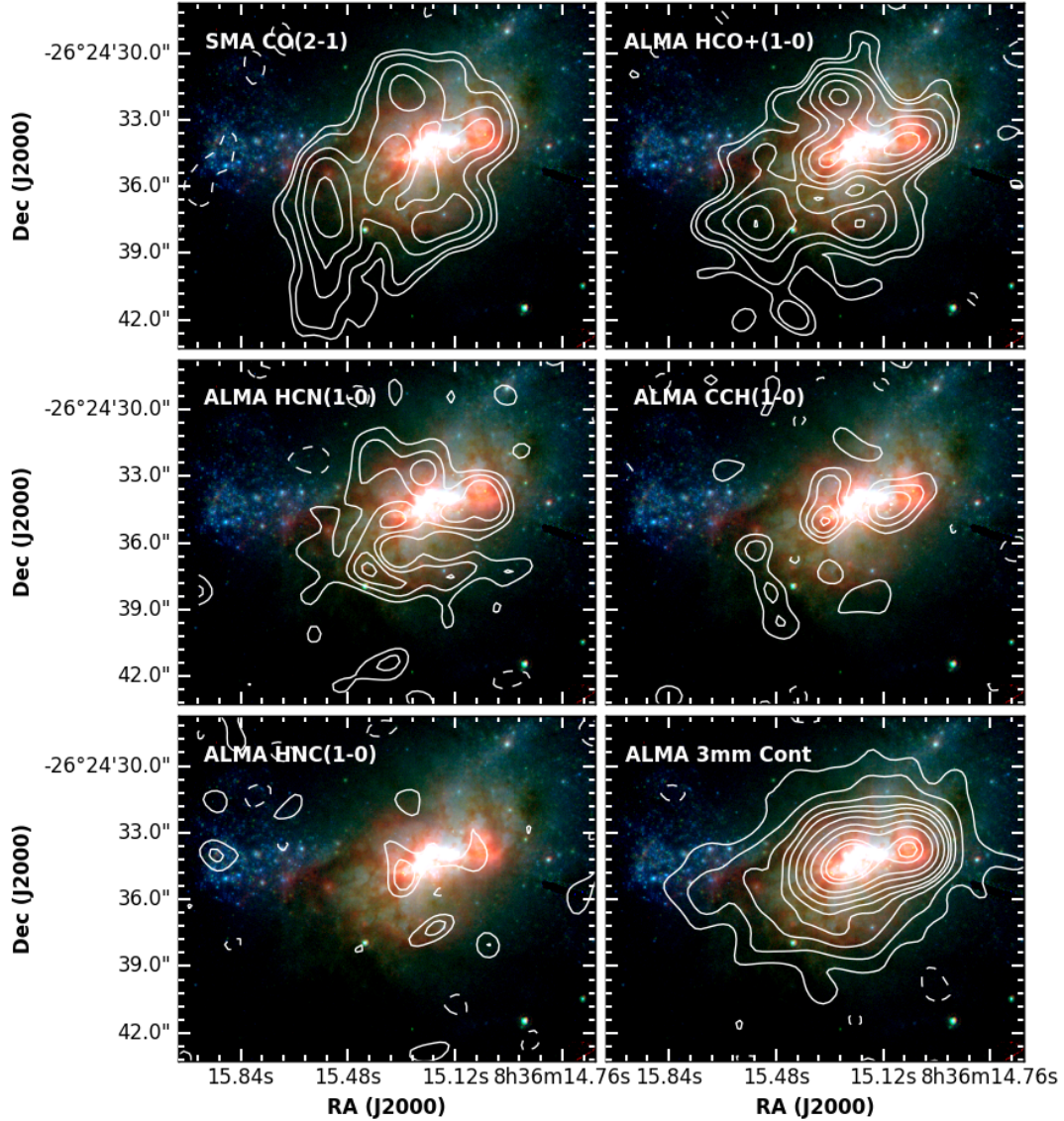}
\caption{Comparison of the molecular emission in He~2-10 observing
  with ALMA in Cycle~0 and with the SMA \citep{santangelo09}.  The
  underlying color image is from HST: red=F187N ($\sim$Pa$\alpha$),
  green=F814W ($\sim$I-band), and blue=F439W ($\sim$B-band).  CO(2-1)
  contours are -2, 2, 3, 5, 7, 10$\sigma$ ($\sigma=1.9$Jy/beam km/s).
  HCO$^+$(1-0) contours are -2, 2, 3, 5, 7, 10,13$\sigma$
  ($\sigma=1.9\times 10^{-2}$Jy/beam km/s).  HCN(1-0) and
    HNC(1-0) contours are 2, 3, 4, 5$\sigma$ ($\sigma=2.0\times
  10^{-2}$Jy/beam km/s).  CCH(1-0) contours are -2, 2, 3, 4, 5$\sigma$
  ($\sigma=2.0\times 10^{-2}$Jy/beam km/s).  Continuum contours are -2, 2,
  6, 12, 18, 24, 36, 60, 90, 150, and 180$\sigma$ ($\sigma=1.1\times
  10^{-5}$Jy/beam).  Negative contours are indicated with dashed lines.
\label{4panel}}
\end{figure*}

\begin{figure*}
\centering
\includegraphics[width=6in]{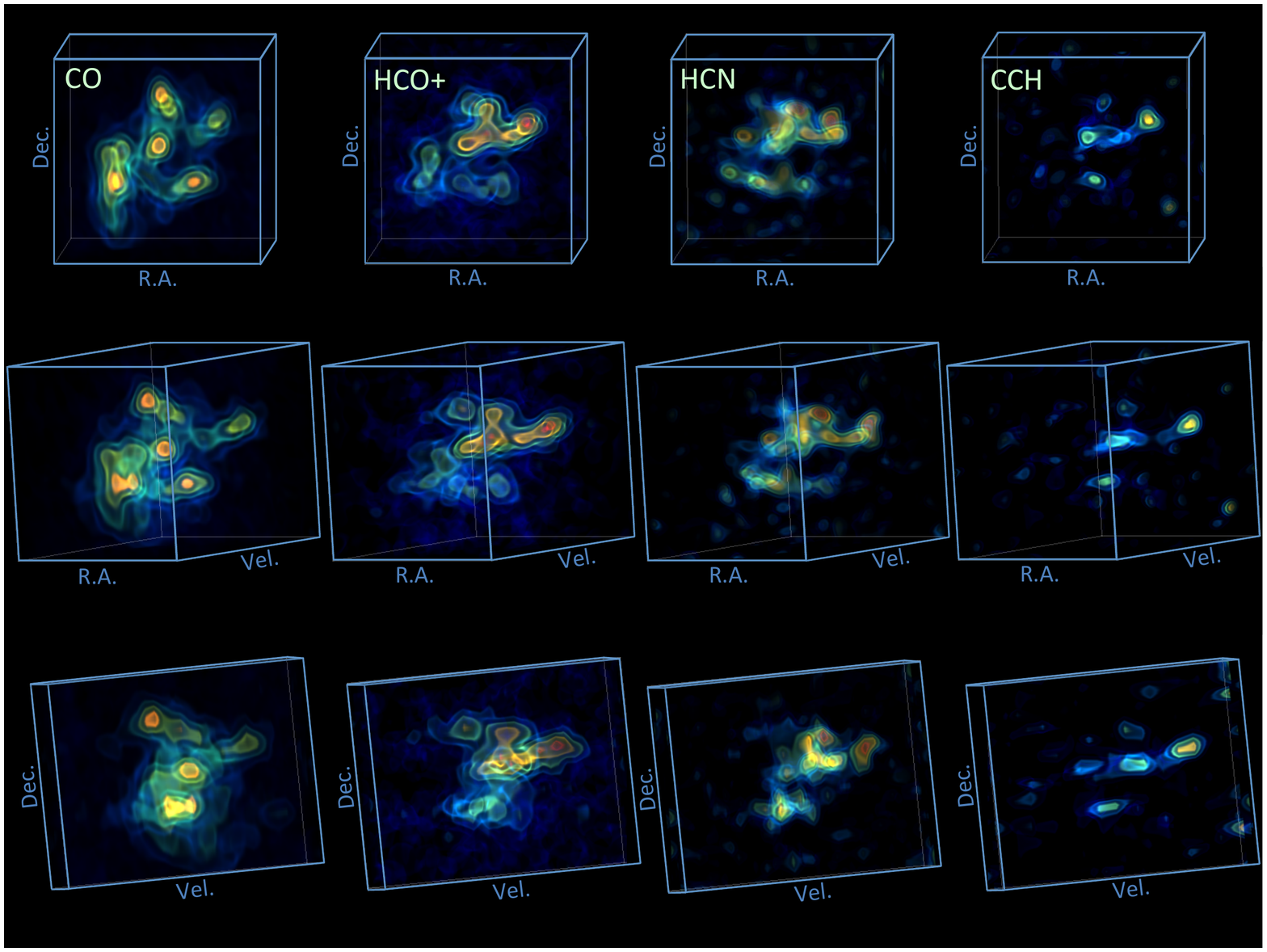}
\caption{Rotations of the velocity cubes for the CO(2-1), HCO$^+$(1-0),
  HCN(1-0), and CCH(1-0) data presented in this paper, with each
  column corresponded to a given line labeled in the top row.  These
  projections were created using the yt program \citep{turk11}, and
  correspond to image cubes with a velocity range of
  770 -- 930~km~s$^{-1}$, an R.A. range of 08:36:15.8 -- 08:36:14.7, and
  an declination range of -26:24:42.7 -- -26:24:27.5.
\label{vel_rot}}
\end{figure*}

Henize~2-10 was observed with ALMA Band~3 (86.9-90.7~GHz and
98.8-102.5~GHz) between May and December of 2012 (program code
2011.0.00348.S). The total time on source was 3.6~hours, with the
number of antennas on-line varying between 15-25 and projected
baseline lengths from $\sim 50-400$~m. Titan and Ceres were used as
the primary amplitude calibrators, and based on the observed variation
we estimate the flux uncertainty to be $\sim 15$\%.  The bandpass
calibrator used was J053851-440507, and the phase calibrator was
J082601-223027.  Synthesis images were created with Briggs weighting
and a robust parameter of 0.5.  Spectral regions around the emission
lines were flagged and the remaining 7GHz of bandwidth in 4 spectral
windows was combined to create a 3.3~mm continuum image with angular
resolution of $1.''6 \times 1.''5$ (70~pc$\times$65~pc) and rms of
1.1$\times$10$^{-5}$~Jy$\;$bm$^{-1}$.  The HCN(1-0), HNC(1-0),
HCO$^+$(1-0), and CCH(1-0) lines were imaged at $1.''7 \times 1.''6$
(75~pc$\times$70~pc) resolution with an rms of 0.45~mJy$\;$bm$^{-1}$
per 10$\;$km$\;$s$^{-1}$ channel.  All lines were strongly detected
(Figure~\ref{4panel}).  Rotations of the full velocity cubes created
with the {\em yt} program \citep{turk11} are shown in
Figure~\ref{vel_rot}.  The telescope primary beam (55$''$) is much
larger than the galaxy (10$''$) so no primary beam correction is
required.

We also incorporate existing SMA CO(2-1) observations with a
resolution of $1.''9 \times 1.''3$ \citep{santangelo09} in this
analysis.  SMA data was combined from both a compact and an extended
  configuration, resulting in baseline lengths between 10 and 180m,
  and imaged with natural weighing.  The rms noise for these
observations was 19~mJy~beam$^{-1}$ per 5~km~s$^{-1}$ channel.  The original
observations and data reduction are described in \citet{santangelo09}.

\begin{figure*}
\centering
\includegraphics[width=5in]{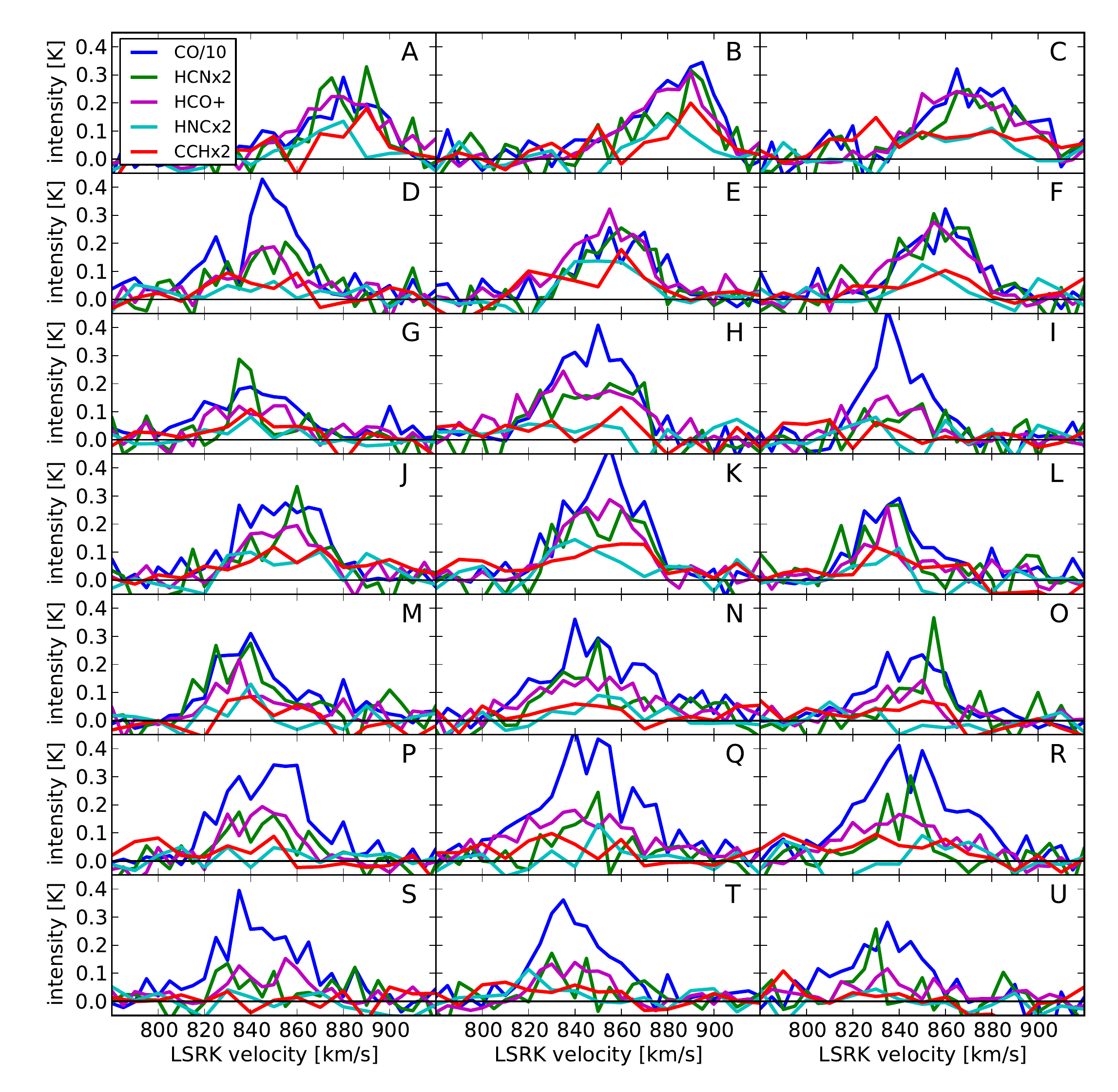}
\caption{The spectra for each of the molecular species used in this analysis.  The regions A-U discussed in Section~\ref{Line_Ratios} are annotated.  To facilitate a comparison on the same vertical scale, the strength of the CO(2-1) line has been decreased by a factor of 10, and the HCN(1-0), HNC(1-0), and CCH(1-0) lines have been increased by a factor of 2.  
\label{spectral_lines}}
\end{figure*}

The spectra of molecular species discussed here for each of the
regions analyzed in this paper are shown in
Figure~\ref{spectral_lines}.  Total intensity maps (moment~0)
  were created by integrating between velocities of
  770-935~km~s$^{-1}$.  Peak intensity (moment~8) maps were created
  using the following thresholds: 160~mJy~beam$^{-1}$ for CO(2-1),
  2.4~mJy~beam$^{-1}$ for HCO$^+$(1-0), 2.0~mJy~beam$^{-1}$ for HCN(1-0)
  and HNC(1-0), and 1.5~mJy~beam$^{-1}$ for CCH(1-0).

\section{Results}
 \label{results}

\subsection{Morphology of the molecular emission \label{morphology}}
The different molecular lines observed here have the potential to aid
in the physical interpretation of the starburst taking place in
He~2-10.  To first order their critical densities (n$_{crit}$) and
transition energies can provide insight into their physical
environment, even if the full astrochemical solutions including
formation and destruction pathways are undoubtedly more complex.

\begin{figure*}[t]
\centering
\plotone{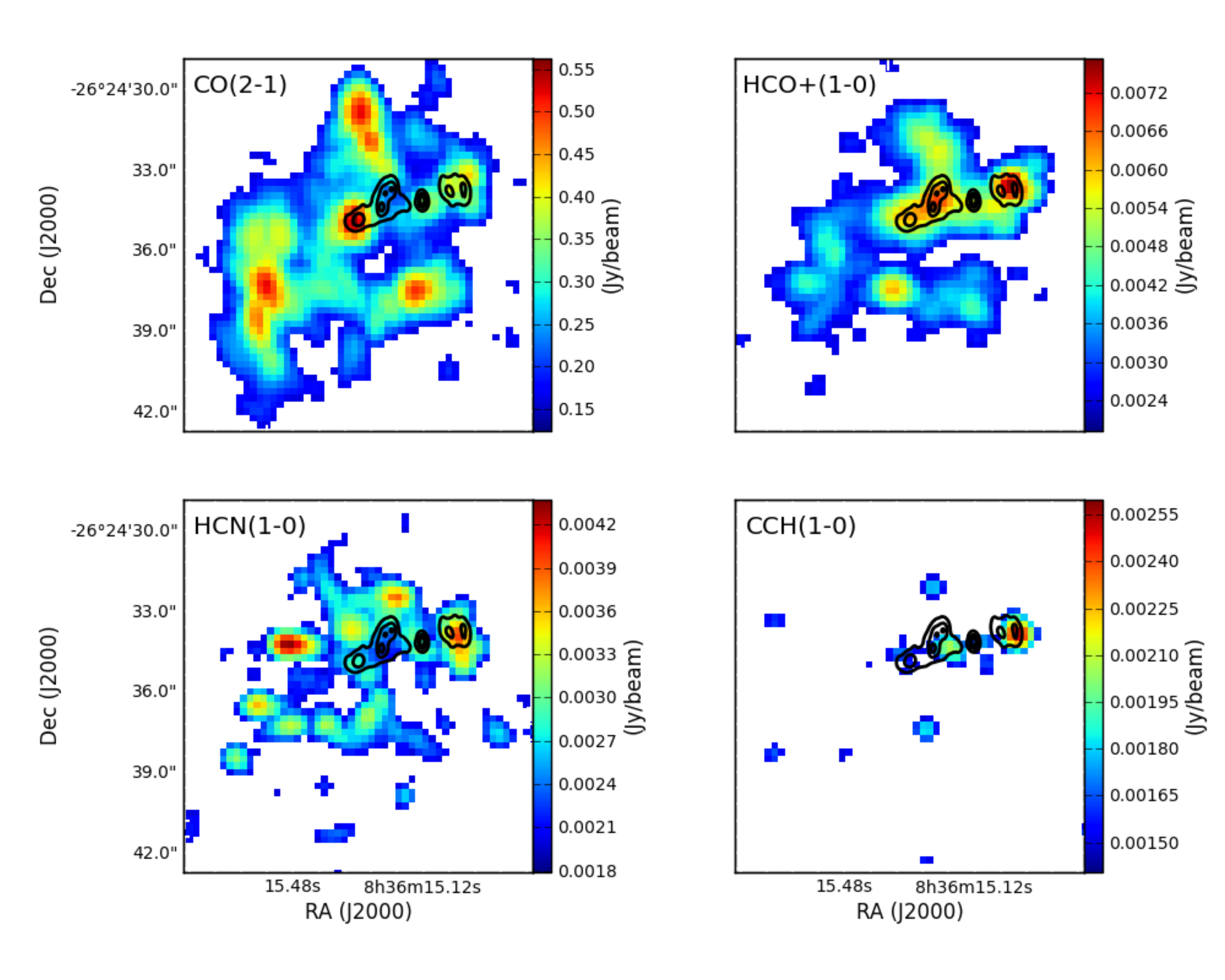}
\caption{Peak brightness intensity (moment 8) maps of CO(2-1),
  HCO$^+$(1-0), HCN(1-0), and CCH(1-0) overlaid with 3.6~cm contours
  indicating thermal radio sources (with the exception of the central AGN.)  }
\label{mom8}
\end{figure*}

\begin{figure}
\centering
\plotone{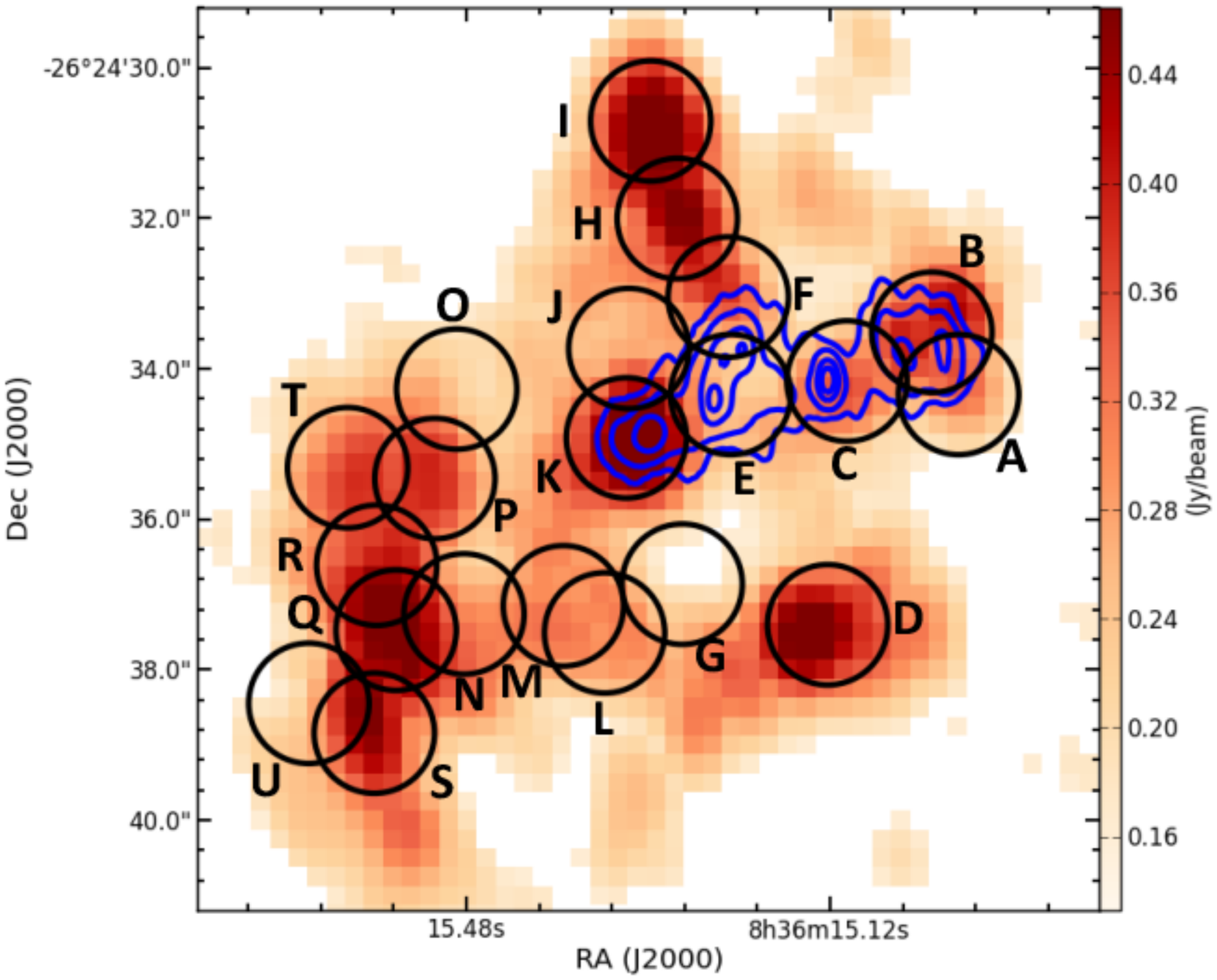}
\caption{Color-scale of the Moment~8 (peak intensity) map of CO(2-1),
  with 3.6~cm contours overlaid.  The regions indicate peaks in one or
  more of the Moment~8 maps of CO(2-1), HCO$^+$(1-0), or HCN(1-0) (see
  Section~\ref{Line_Ratios}).  The regions have a radius of $0.''8$,
  and are approximately the size of the beams.
}
\label{Mom8_regions}
\end{figure}

To facilitate the analysis of these observations we define regions of
$0.''8$ radius (roughly the beam size) around peaks in the Moment~8
(peak intensity) maps for each transition, as shown in
Figure~\ref{mom8} and Figure~\ref{Mom8_regions}.  Fluxes were
  measured in r=0.8$''$ circular apertures on the image integrated in
  velocity from 770 to 935 km$\;$s$^{-1}$.  The emission lines do not
  peak at precisely the same positions, so to calculate a conservative
  uncertainty, the peak position of each line in each aperture was
  determined, a new flux measured for each recentered aperture, and
  the dispersions of those recentered fluxes used as a measure of the
  uncertainty due to source position.  To that in quadrature was added
  a 20\% absolute calibration uncertainty and the rms noise in images.
  For non detections, 5 times the rms noise of the image is quoted as
  an upper limit.  These values along with the inferred molecular
masses for these regions, M$_{mol}$, are provided in
Table~\ref{line_strengths}.

\begin{center}
\begin{deluxetable*}{lllcccccc}[t]
\tabletypesize{\footnotesize}
\tablecolumns{8}
\tablewidth{0pc}
\tablecaption{Line Strengths}
\tablehead{
\colhead{ID} & \colhead{R.A.}    & \colhead{Dec.} &
\colhead{CO(2-1)}   &\colhead{HCO$^+$(1-0)}    & \colhead{HCN(1-0)} &
\colhead{HNC(1-0)} &  \colhead{CCH(1-0)} & \colhead{M$_{mol}$} \\
\colhead{} & \colhead{(J2000)}    & \colhead{(J2000)} & \colhead{(K km s$^{-1}$)}   &\colhead{(K km s$^{-1}$)}  &\colhead{(K km s$^{-1}$)} &\colhead{(K km s$^{-1}$)} & \colhead{(K km s$^{-1}$)}  & \colhead{($10^6$~M$_{\odot}$)}}
\startdata
A&	8h36m14.99s&-26d24m34.4s&	87.6 $\pm$17.5&	9.0$\pm$1.8&  4.8$\pm$1.0 & $<1.2$ & 2.9 $\pm$ 0.7&	 3.5 \\
B&	8h36m15.02s&-26d24m33.5s&	137.3$\pm$27.5&	10.3$\pm$2.1&4.2$\pm$0.8 & 1.4 $\pm$ 0.5 & 3.4 $\pm$ 1.0&	 5.5 \\
C&	8h36m15.10s&-26d24m34.2s&	142.9$\pm$28.6&  12.5$\pm$2.5&4.2$\pm$0.8 & 2.1 $\pm$ 0.4 & 5.5 $\pm$ 1.1&  5.8 \\
D&	8h36m15.12s&-26d24m37.4s&	157.9$\pm$31.6&	6.8$\pm$1.4&3.9$\pm$0.8 & 1.5 $\pm$ 0.7 & 2.3 $\pm$ 0.5&	 6.5 \\
E&	8h36m15.22s&-26d24m34.4s&	102.4$\pm$20.5&12.3$\pm$2.5&3.6$\pm$0.7& 2.8 $\pm$ 0.8 & 1.4 $\pm$ 0.6& 4.1 \\
F&	8h36m15.22s&-26d24m33.1s&	114.4$\pm$22.9&	10.1$\pm$2.0&4.6$\pm$0.9 & 2.0 $\pm$ 0.7 & 1.4 $\pm$ 0.3& 1.9 \\
G&	8h36m15.27s&-26d24m36.9s&	106.7$\pm$21.3&	5.6$\pm$1.1&2.0$\pm$0.4 & 1.3 $\pm$ 0.5 & 1.3 $\pm$ 0.4&	4.6 \\
H&	8h36m15.27s&-26d24m32.0s&	148.3$\pm$29.7& 11.8$\pm$2.4&3.1$\pm$0.6 & 1.9 $\pm$ 0.5 & 1.1 $\pm$ 0.3& 5.9 \\
I&	8h36m15.30s&-26d24m30.7s&	107.1$\pm$21.4&	4.8$\pm$1.0&1.8$\pm$0.4& $<1.2$ & 1.5 $\pm$ 0.3&	4.3 \\
J&	8h36m15.32s&-26d24m33.7s&	141.3$\pm$28.3&	7.9$\pm$1.6&2.4$\pm$0.5 & 2.4 $\pm$ 0.6 & 3.8 $\pm$ 1.0&	5.7 \\
K&	8h36m15.32s&-26d24m34.9s&	169.0$\pm$33.8 &11.5$\pm$2.3&4.3$\pm$0.9 & 2.8 $\pm$ 0.6 & 3.7 $\pm$ 0.9& 6.8 \\
L&	8h36m15.34s&-26d24m37.5s&	107.1$\pm$21.4&	5.4$\pm$1.1&4.0$\pm$0.8& $<1.2$ & 1.4 $\pm$ 0.7&	4.3 \\
M&	8h36m15.38s&-26d24m37.2s&	114.3$\pm$22.9&	5.4$\pm$1.1&4.9$\pm$1.0 & $<1.2$ & $<1.1$&	4.6 \\
N&	8h36m15.48s&-26d24m37.3s&	163.7$\pm$32.7&	9.0$\pm$1.8&3.4$\pm$0.7 & $<1.2$ & 2.2 $\pm$ 0.5&	6.5 \\
O&	8h36m15.49s&-26d24m34.3s&	85.7$\pm$17.1&	4.3$\pm$0.9&2.3$\pm$0.5 & $<1.2$ & 1.5 $\pm$ 0.3&	3.4 \\
P&	8h36m15.51s&-26d24m35.5s&	153.1$\pm$30.6&	4.9$\pm$1.0&2.8$\pm$0.6 & 1.4 $\pm$ 0.6 & 1.5 $\pm$ 0.4&	6.1 \\
Q&	8h36m15.55s&-26d24m37.5s&	220.2$\pm$44.0&11.1$\pm$2.2&2.4$\pm$0.5 & $<1.2$ & 1.7 $\pm$ 0.5&8.8 \\
R&	8h36m15.57s&-26d24m36.6s&	205.6$\pm$41.1&	7.6$\pm$1.5&2.8$\pm$0.6 & $<1.2$ & 1.5 $\pm$ 0.5&	8.2 \\
S&	8h36m15.57s&-26d24m38.9s&	134.2$\pm$26.8&  5.0$\pm$1.0&1.3$\pm$0.3 & $<1.2$ & $<1.1$&	 5.4 \\
T&	8h36m15.60s&-26d24m35.3s&	131.4$\pm$26.3&	3.8$\pm$0.8&2.4$\pm$0.5 & 1.2 $\pm$ 0.3 & 1.2 $\pm$ 0.3&	 5.3 \\
U&	8h36m15.64s&-26d24m38.4s&	87.5$\pm$17.5&	4.1$\pm$0.8&1.3$\pm$0.3 & $<1.2$ & $<1.1$&	3.5\\
\enddata
\tablecomments{M$_{mol}$ calculated assuming $\alpha_{CO} = 4.3$ (K km s$^{-1}$
pc$^2$)$^{-1}$ for regions with a radius of $0.''8$ ($=35$pc at a
distance of 9~Mpc).}
\label{line_strengths}
\end{deluxetable*}
\end{center}

\subsubsection{CO(2-1) Emission}
Given its high abundance and low critical density ($\sim 10^3$~cm$^{-3}$),
CO is an important ``generic'' molecular gas tracer.  This line has the
highest upper state energy of the tracers used here of T$=16.6$~K.
CO is also relatively easily photo-dissociated from the ground-state,
requiring only 11.09~eV.  Thus, we expect CO to be rapidly destroyed
amid the strong ultraviolet emission from young massive stars.

While CO(2-1) emission is detectable throughout the body of He~2-10,
it is not preferentially peaked near the locations of the natal super
star clusters (see Figure~\ref{mom8}).  In fact, only the eastern-most
thermal radio source shows particularly strong CO(2-1) emission
(region~K, see Fig.~\ref{Mom8_regions}), suggesting that perhaps this
cluster is among the youngest in the system.  However, a number of
distinct CO clouds are located throughout the full 3D cube.
Assuming a CO line intensity to mass conversion factor of $\alpha_{CO}
= 4.3$~M$_\odot$~pc$^{-2}$~(K~km~s$^{-1}$)$^{-1}$ \citep{bolatto13}
and adopting a conversion of CO(2-1)$= 0.5\times$CO(1-0) these
$0.''8$ regions have masses of up to $\sim 9\times
10^6$~M$_{\odot}$ and likely indicate the regions where the next
generation of star clusters will form.

\subsubsection{HCN(1-0) and HNC(1-0) Emission}
HCN(1-0) is the highest critical density tracer used here, with
$n_{crit}=1.7\times 10^5$~cm$^{-3}$ at 50~K for optically thin gas
\citep{shirley15}\footnote{Note that for optically thick gas,
  radiative trapping will result in an ``effective'' excitation
  density which can be significantly lower than the value calculated
  for optically thin gas \citep{shirley15}}, but has an
upper state energy of T$=4.3$~K, similar to HCO$^+$.  HNC(1-0) has
a somewhat lower critical density than HCN(1-0) with
$n_{crit}=8.4\times 10^4$~cm$^{-3}$ at 50~K for optically thin gas
\citep{shirley15}, but requires an upper state energy of
T$=4.4$~K.

The relative strength of these isomers is likely driven by the
hydrogen exchange reaction $H+CNH \leftrightarrows HCN+H$
\citep{talbi96}.  The activation barrier for these two reactions is
quite different, with the $HCN+H\rightarrow HNC+H$ requiring four
times the activation energy of $HNC+H\rightarrow HCN+H$
\citep{talbi96, graninger14}.  The behavior of the HNC/HCN ratio was
studied extensively by \citet{schilke92} and appears to show a strong
inverse temperature dependence -- with HNC/HCN near unity in the cold
regions of OMC-1, but falling to $\sim 1/80$ in the warm Orion-KL
region.

For these ALMA observations of He~2-10, we would have predicted
relatively higher molecular gas temperatures in the vicinity of the
natal super star clusters.  Although the HNC is only detected at
  the 3-5$\sigma$ level, and only in the vicinity of the nascent
  SSCs, the apparent HNC/HCN ratio in these areas varies
between $\sim 0.3 - 1$.  If we
assume this ratio is inversely related to temperature, this suggests
the warmest molecular regions are in the vicinity of regions B and F,
which are in the vicinity of natal super star clusters detected via
thermal radio emission.  However, regions E and K which are also
associated with young super star clusters have ratios of HNC/HCN$>0.5$
-- indicating cooler temperatures.  Thus, our original hypothesis that
the HNC/HCN ratio would be lowest near the natal clusters in not
confirmed.  However, it is entirely possible that given the spatial
resolution of these data, each region could contain a range of
physical conditions, and some gas affected more and some less by the
natal clusters, which complicates the analysis.

\begin{figure*}[t!]
\centering
\plotone{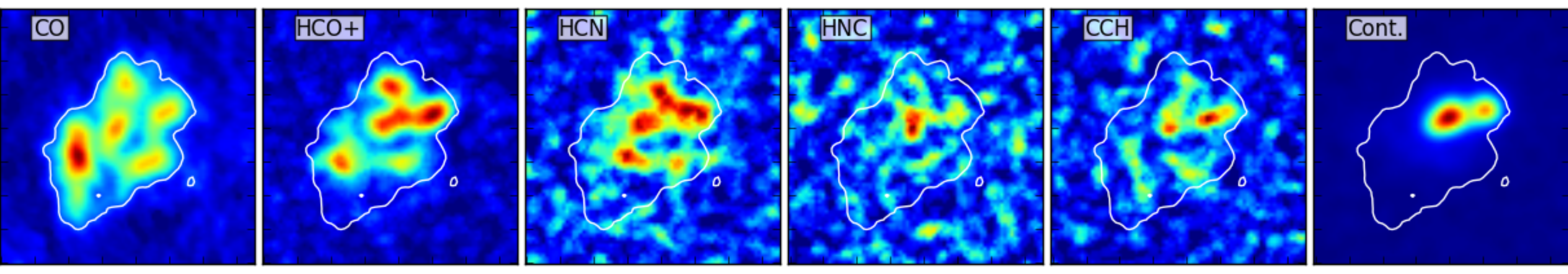}
\plotone{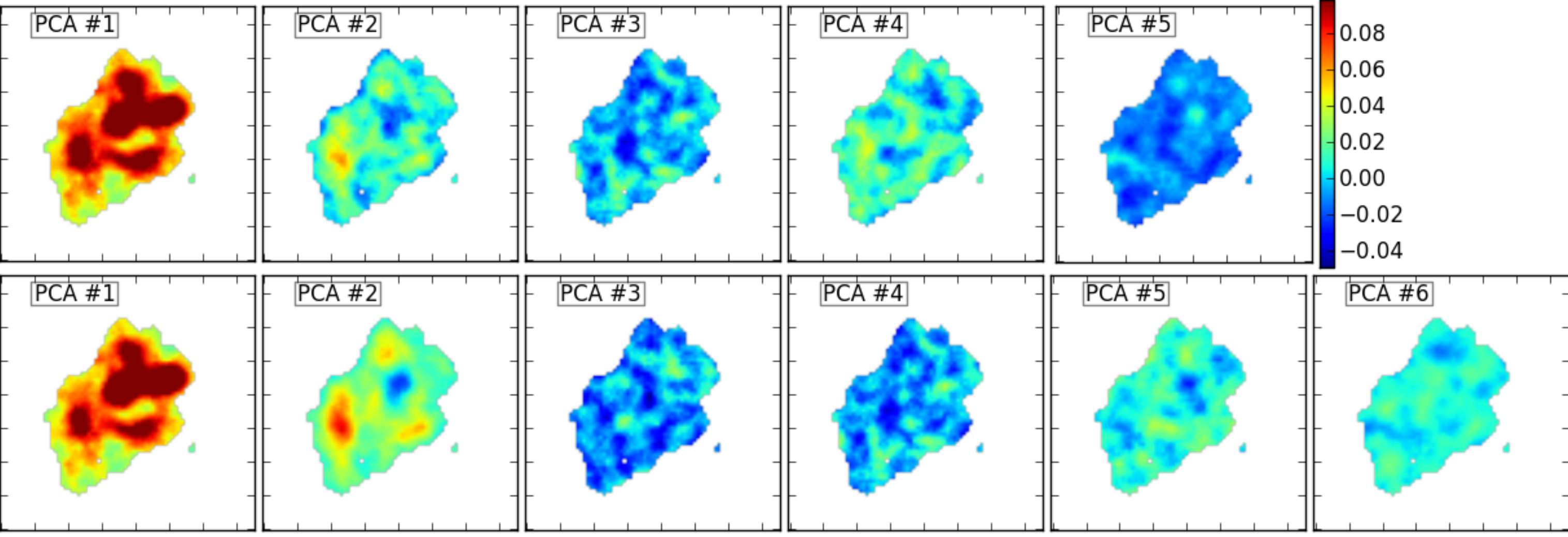}
\caption{(top) Integrated intensity images of the lines used in the
  PCA analysis with a white outline indicating the 3$\sigma$ threshold
  of CO emission for which the analysis was done.  (middle) The
  intensity of the principle component images using line emission
  only.  (bottom) The intensity of the principle component images
  including the 3~mm continuum emission. \label{PCA_int}}
\end{figure*}

\subsubsection{HCO$^+$(1-0) Emission}
With an optically-thin critical density of $n_{crit}=2.9\times
10^4$~cm$^{-3}$ at 50~K \citep{shirley15}, and an upper state energy
of T$=4.3$~K, HCO$^+$ is a prevalent moderate-density tracer.  Of the
molecular lines studied here, HCO$^+$ is the most closely associated
with the natal super star clusters (see Figure~\ref{mom8}), with
strong peaks on each of the sources of thermal radio emission.
Curiously, the HCO$^+$ and HCN emission are not particularly well
correlated (see Figure~\ref{mom8} and Section~\ref{PCA}), despite their
similar critical densities and upper state energies; the ratio of
HCN/HCO$^+$ varies between $\sim 0.2 - 0.9$ across He~2-10, with the
highest values associated with regions outside of the area of thermal
radio emission.

\subsubsection{CCH(1-0) Emission}
The CCH(1-0) emission observed here is a blend of two hyperfine lines
at 87.317~GHz and 87.329~GHz $((1_{3/2,1}-0_{1/2,0})$ and $(1_{3/2,2}-0_{1/2,1}) )$,
which have upper state energies of T$=4.2$~K.  CCH is one of the
most abundant hydrocarbons and is believed to be enhanced by CO
dissociation in strong ultraviolet radiation fields \citep[e.g.,][]{beuther08,martin14}.  
Therefore, it is not surprising that the CCH emission peaks near the
locations of the natal super star clusters (see Figure~\ref{mom8})
where the CO emission is relatively low.

The CCH may also indicate another interesting feature: as noted in
Section~1, the CO-bright Eastern cloud may be associated with a
gas-rich cloud falling into the galaxy; \citet{kobulnicky95}
  detect a kinematically distinct elongated CO feature to the
  south-east of the main galaxy (with the southern end pointing at
  RA=8h36m15.48s), which could be interpreted as an infalling gas
  cloud.  In Figure~\ref{4panel} a bright elongated region of CCH
emission that can be seen running northeast-southwest.
  This CCH feature is coincident with the inside edge of the
  kinematically distinct CO cloud -- if the CO cloud is falling in,
  the CCH could be highlighting the area of intersection.  This
correspondence could be merely circumstantial, but may indicate the
radiation from the central star-forming zone is leaking out and
illuminating the front edge of the in-falling CO cloud.

\subsection{Line Widths \label{line_widths}}
The width of the emission lines from each of the regions identified in
Figure~\ref{Mom8_regions} were fit with a Gaussian
model and measured in CASA.   In general, the strength of the HNC and
CCH emission was not strong enough to measure a line width, and for a few
regions this was also true of the HCN and/or HCO$^+$ lines.  The
resulting line widths are presented in Table~\ref{sigma_tab}.   

\begin{deluxetable}{lllcccccc}[t]
\tabletypesize{\footnotesize}
\tablecolumns{4}
\tablewidth{0pc}
\tablecaption{Line Widths}
\tablehead{
\colhead{ID} & \colhead{$\sigma$ CO(2-1)} & \colhead{$\sigma$ HCN(1-0)} & \colhead{$\sigma$ HCO$^+$(1-0)} \\ 
\colhead{} & \colhead{(km s$^{-1}$)}   &\colhead{(km s$^{-1}$)}  &\colhead{(km s$^{-1}$)} }
\startdata
A & $17.1 \pm 3.4$ & $14.4 \pm 2.9$ & $18.5\pm 3.7$ \\
B & $15.6 \pm 3.1$ & $15.6 \pm 3.1$ & $16.6\pm 3.3$ \\
C & $20.4 \pm 4.1$ & $19.2 \pm 3.8$ & $20.5\pm 4.1$  \\
D & $15.0 \pm 3.0$ & $17.7 \pm 3.5$ & $14.4\pm 2.9$ \\
E & $17.4 \pm 3.5$ & $13.3 \pm 2.7$ & $17.1\pm 3.4$ \\
F & $15.7 \pm 3.1$ & $14.8 \pm 3.0$ & $15.8\pm 3.2$  \\
G & $21.8 \pm 4.4$ & ---          & ---   \\
H & $16.3 \pm 3.3$ & $17.4 \pm 3.5$ & $24.7\pm 4.9$   \\
I & $12.3 \pm 2.5$ & ---          & $14.8\pm 3.0$  \\
J & $18.3 \pm 3.7$ & $13.4 \pm 2.7$ & $14.8\pm 3.0$ \\
K & $16.1 \pm 3.2$ & $17.7 \pm 3.5$ & $15.3\pm 3.1$  \\
L & $14.2 \pm 2.8$ & $12.7 \pm 2.5$ & $10.6\pm 2.1$  \\
M & $15.3 \pm 3.1$ & $14.5 \pm 2.9$ & $12.6\pm 2.5$ \\
N & $24.2 \pm 4.8$ & $14.3 \pm 2.9$ & $25.3\pm 5.1$  \\
O & $16.0 \pm 3.2$ & ---          & $14.6\pm 2.9$  \\
P & $18.3 \pm 3.7$ & $16.2 \pm 3.2$ & $14.8\pm 3.0$ \\
Q & $21.6 \pm 4.3$ & ---          & $27.2\pm 5.4$ \\
R & $22.4 \pm 4.5$ & ---          & $21.4\pm 4.3$  \\
S & $19.5 \pm 3.9$ & ---          & $14.9\pm 3.0$ \\
T & $14.6 \pm 2.9$ & ---          & $14.6\pm 2.9$  \\
U & $17.6 \pm 3.5$ & ---          & --- 
\enddata
\label{sigma_tab}
\end{deluxetable}

\begin{figure*}
\plottwo{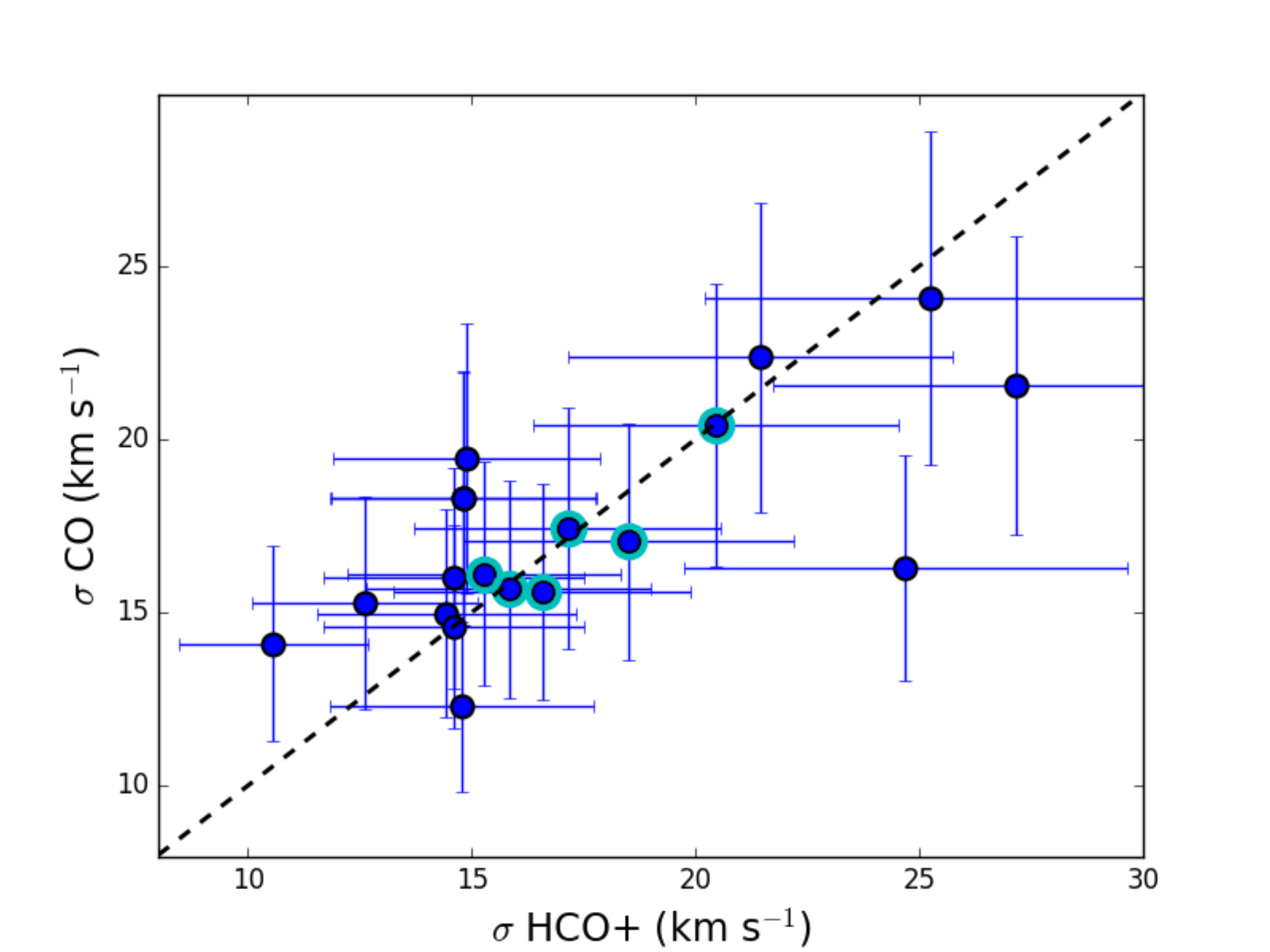}{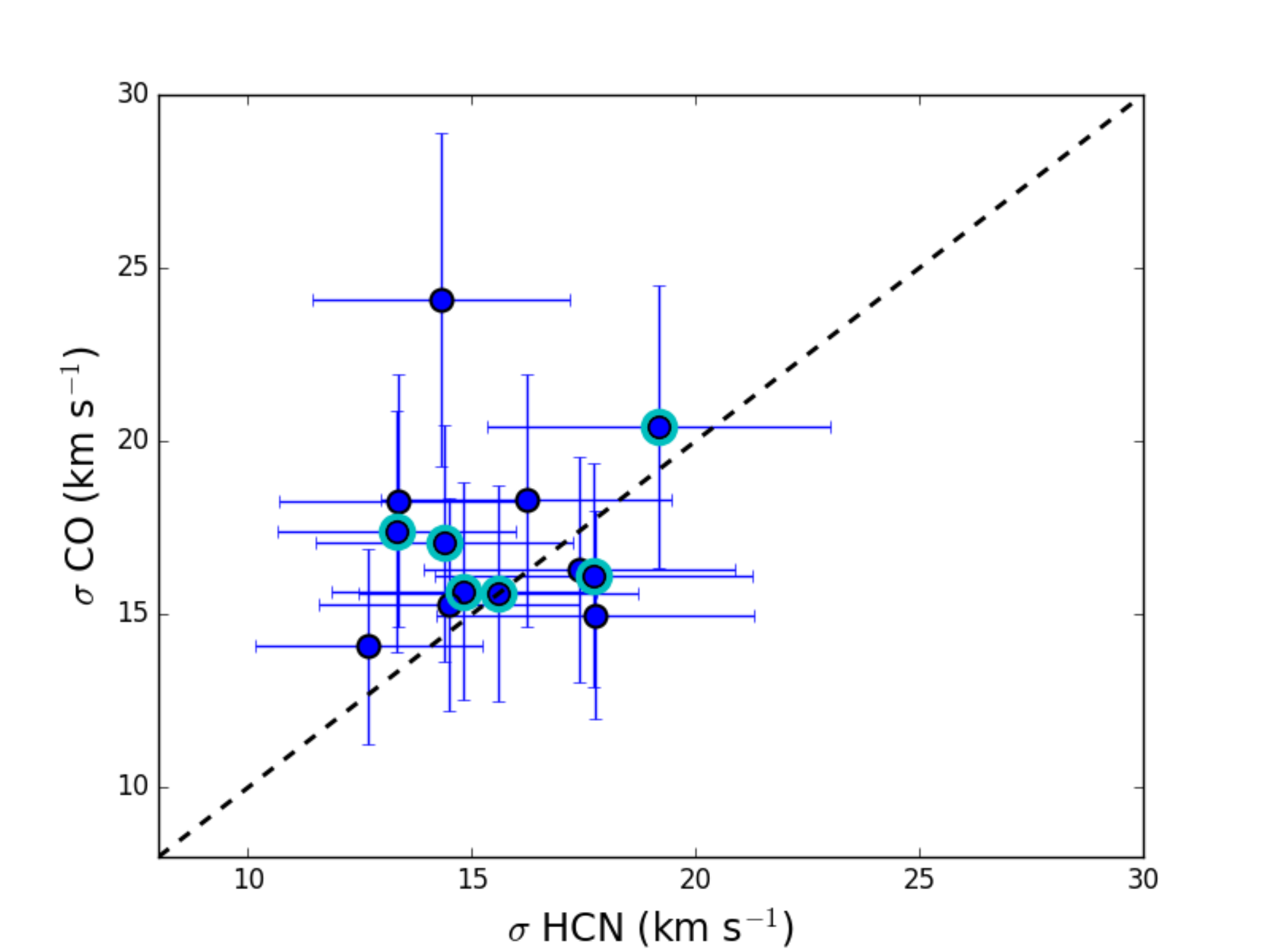}
\caption{The relationship between line widths observed for
    different molecular species. (left) A comparison of the line
    widths of HCO$^+$(1-0) and CO(2-1). (right) A comparison of the
    line widths of HCN(1-0) and CO(2-1).  The dashed line  in
    each panel indicates equivalence.  Regions highlighted in cyan
  include areas associated with thermal radio emission.}
\label{plot_sigmas}
\end{figure*}

As shown in Figure~\ref{plot_sigmas}, the line-widths of CO, HCN, and
HCO$^+$ roughly follow a one-to-one relationship with some scatter.
The lack of an apparent trend in line-width with respect to species is
circumstantial evidence that these molecules are tracking the same
components of a given molecular cloud.  If, for example, any of these
species were preferentially associated with outflows or cloud
collapse, one might predict a difference in the observed line widths.
However, on the size scales of the regions measured here (1.$''$6
diameter, $\sim 70$~pc ), any trends in molecular line widths may well
be rendered invisible by averaging over numerous discrete clouds.  The
regions that include areas of thermal radio emission are also
indicated in Figure~\ref{plot_sigmas}.  We originally hypothesized
that regions containing natal SSCs might preferentially exhibit larger
line-widths as the stars inject energy into the surrounding ISM;
  in so far as line-widths are related to the temperature and
  turbulence in a cloud, these should increase as stars inject energy
  into their surrounding medium.  However there are no apparent
trends with respect to whether a region is associated with thermal
radio emission.

\begin{figure}
\plotone{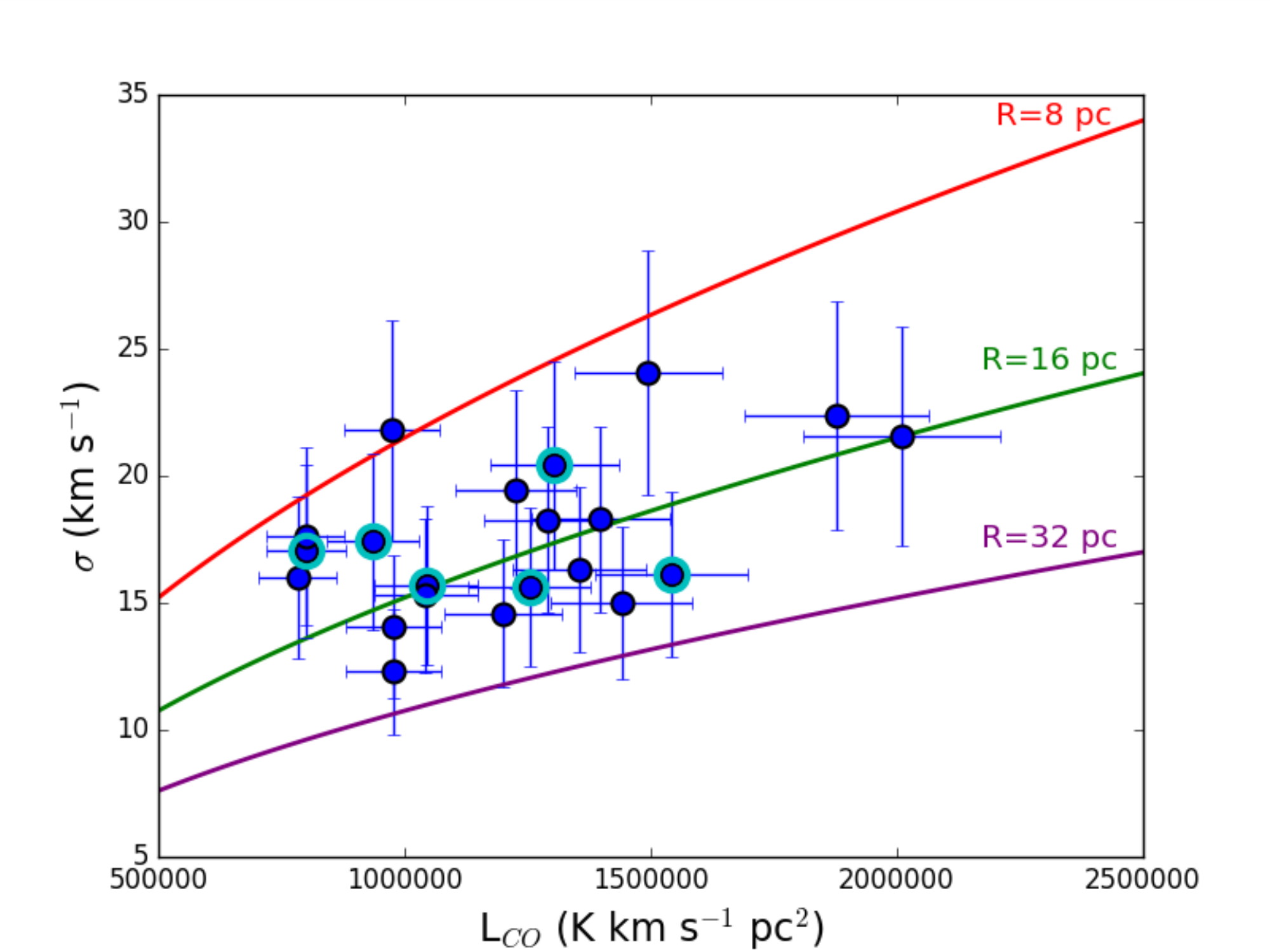}
\caption{The relationship between the CO luminosity and line width.  Regions
  highlighted in cyan include areas with thermal radio emission. The
  red, green, and purple lines indicate where molecular clouds of a
  given size would fall based on \citet{heyer09}.}
\label{plot_COvsCO_sig}
\end{figure}

In addition, we investigate the relationship between the CO luminosity
and the associated line width, as shown in
Figure~\ref{plot_COvsCO_sig}.  Given that these molecular clouds are
not resolved, we cannot put these data points on a size-line-width
relation.  However, based on the work of \citet{heyer09}, we rearrange
the standard size-linewidth-surface density relationship as follows:
\begin{equation}
\sigma_{\rm V}\, = \left(\frac{\alpha_{CO}\ L_{CO}\ G}{\rm 5\ R}\right)^{1/2}.
\end{equation}
We again adopt $\alpha_{CO} =
4.3$~M$_{\odot}$~pc$^{-2}$~(K~km~s$^{-1}$)$^{-1}$ \citep{bolatto13},
and assume CO(2-1)$= 0.5\times$CO(1-0).   Thus, for a given radius, it
is possible to determine the predicted line width based on this
equation.  Lines corresponding to radii of 8~pc, 16~pc, and 32~pc are
over-plotted in Figure~\ref{plot_COvsCO_sig}.   The location of the
data points from the regions in He~2-10 suggest that either 1) the molecular
clouds have radii of $\sim 8-32$~pc, or alternatively 2) the clouds could
be larger, but have enhanced line widths -- potentially due to high
pressure environments as seen in \citet{johnson15}.  

\subsection{Principle Component Analysis}
 \label{PCA}

\begin{figure}[b]
\centering
\plotone{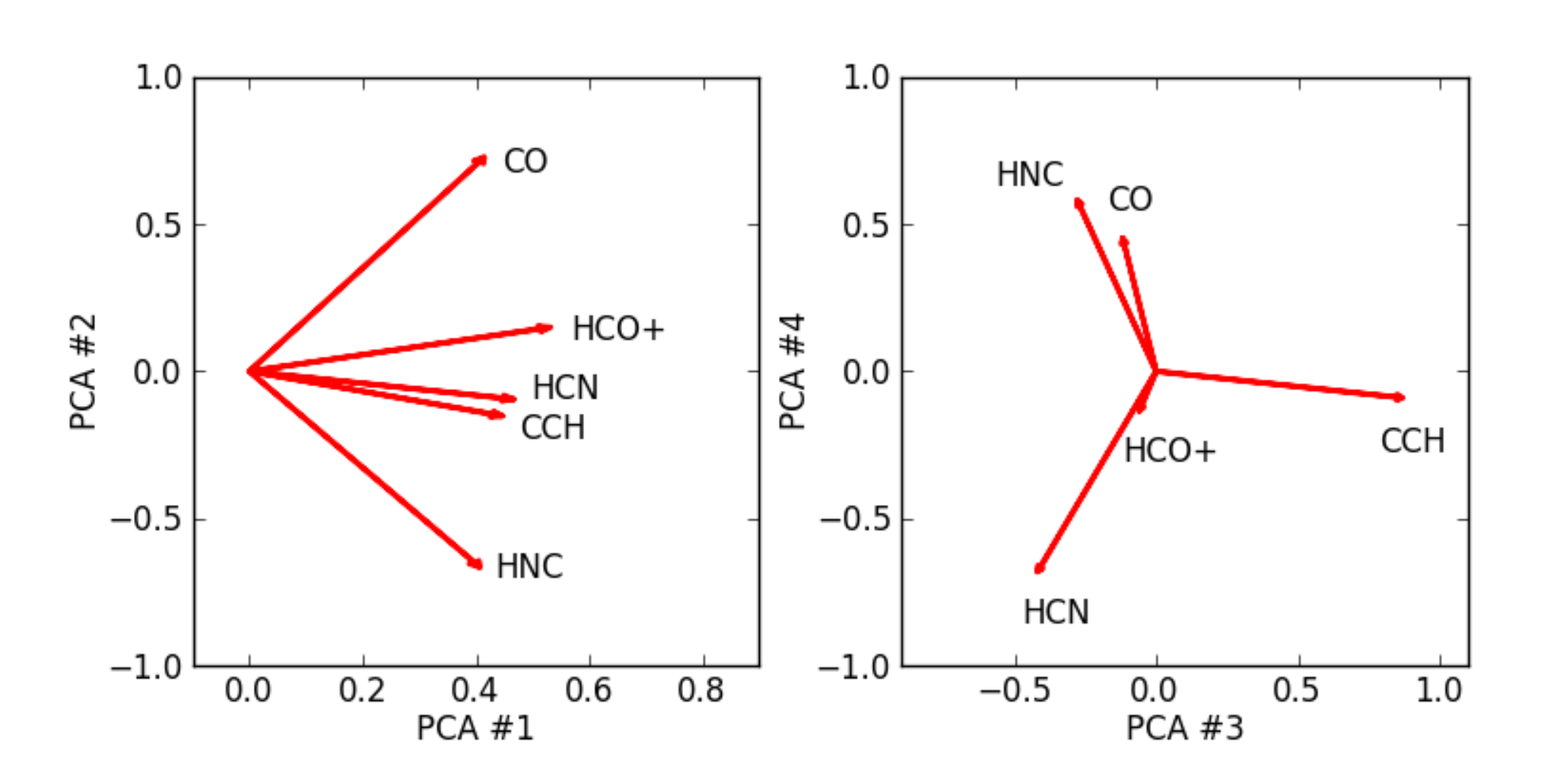}
\plotone{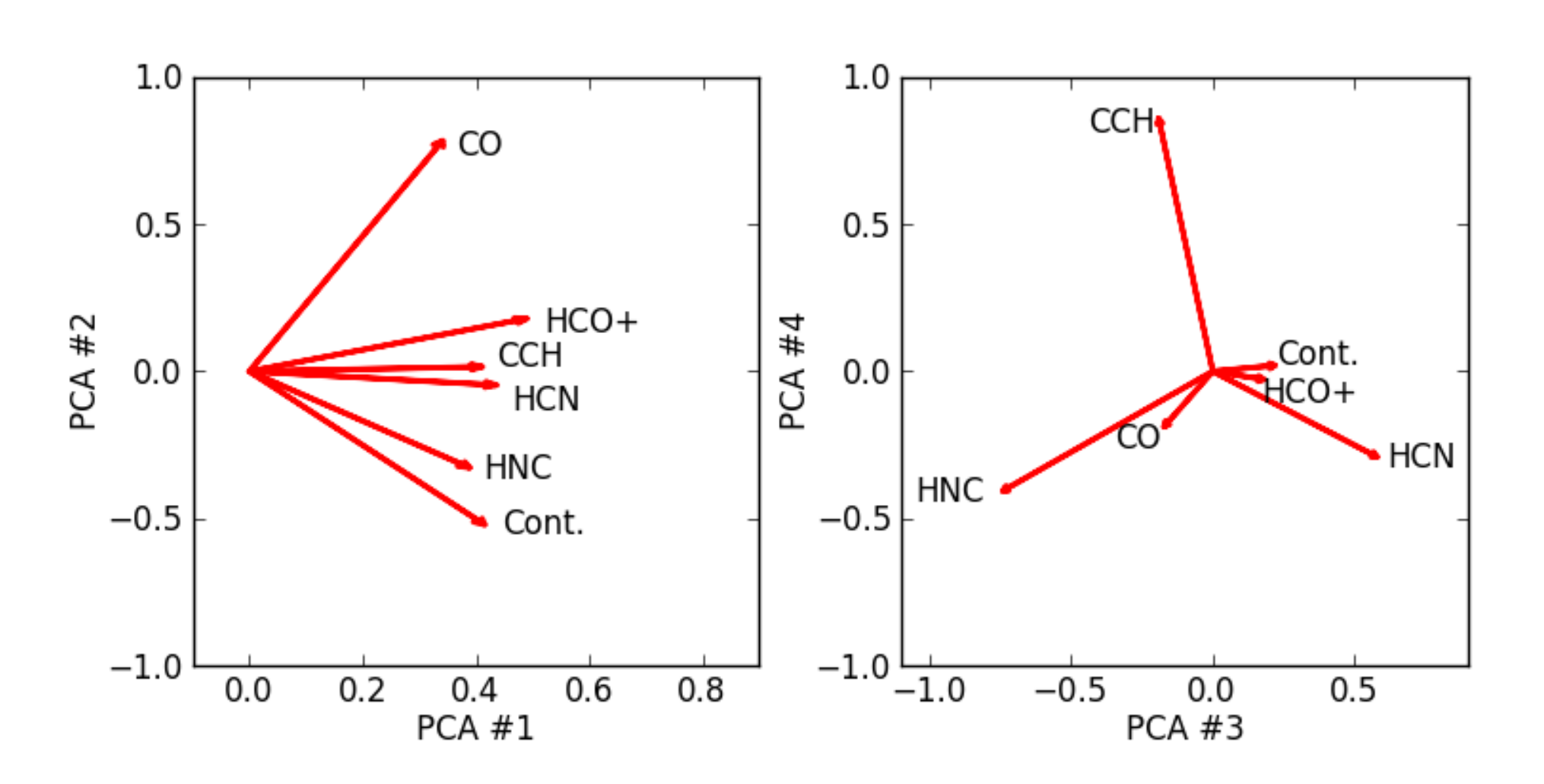}
\caption{Eigenvectors of the PCA analysis. The top two panels show the
  results using emission lines only. The bottom two panels include the
  3~mm continuum emission in the analysis.  PCA\#1 indicates the
  average emission morphology of the galaxy, with the most generic
  tracers being more horizontal in the plot. PCA\#2 indicates the
  extent to which emission from a given line deviates from the PCA\#1
  averaged over the region outlined in white in Fig.~\ref{PCA_int}. In
  the analysis of only lines, PCA\#3 captures the different
  distribution of CCH from the other lines.  This difference is
  captured by PCA\#4 when continuum is included. \label{PCA_vect}}
\end{figure}

To quantify the emission distribution in an objective way, we perform
a Principle Component Analysis (PCA) of the integrated intensity
images.  The images are first normalized before calculation of
eigenvectors and eigenimages in order to avoid the calculation
dominated by the strongest emission (CO in this case).  We tested
  normalizing both each integrated intensity image by its peak, and
  alternatively normalizing each image by its rms. The ratio of peak
  to rms varies between 4.5 and 6.5 and therefore the results do not
change qualitatively depending on the normalization; here we show the
results when images are normalized by their rms.  In order to
  mitigate spurious results, we apply a threshold in the CO(2-1)
  emission of 3$\sigma$, and the associated region is adopted in the
  analysis of all images.  The results are not sensitive to choices
  for that threshold between 2.5 and 4$\sigma$.
Figure~\ref{PCA_int} shows the
integrated intensity images and the principle component images on the
same scale.  Figure~\ref{PCA_vect} shows the eigenvectors.

The PCA analysis reveals several interesting trends.  For example,
looking at the first component (which indicates how similar or
different emission from a given line is to the average emission
morphology of the galaxy); quantitatively, HCO$^+$ is the most generic
tracer of molecular emission (its eigenvector is nearly along PCA\#1
with little projection along the other components in
Figure~\ref{PCA_vect}).
  
The second component (which indicates how different emission
morphology in a given line is from the average) also reveals
interesting features.  From the PCA\#2 vectors (Fig.~\ref{PCA_vect}),
it is apparent that the CO(2-1) and HNC(1-0) are the most different,
and strongly anti-correlated.  The morphology of the PCA\#2 intensity
map (Fig.~\ref{PCA_int}) and its morphology highlights the
southeastern cloud complex (containing sources N,U,Q,R,S,T), which is
particularly strong in CO emission, but has limited emission in the
other lines.  It is also noteworthy that CO and CCH show substantially
different vectors, which would be expected if CCH emission is enhanced
by CO photodissociation.  Curiously the CO and continuum emission are
the most different, suggesting that the either the dust and molecular
gas components of the galaxy are not cospatial, or the 88~Ghz
continuum is dominated by free-free emission (see
Section~\ref{dust_mass}).  A similar result was previously observed by
\citet{kepley16} for the relatively low metallicity galaxy IIZW40
\citep[12+log(O/H)=8.09, $\sim 1/5$~Z$_{\odot}$][]{guseva00}, where
they found that dust emission was not significant below frequencies of
$\sim 100$~Ghz.  Given the metallicity of IIZW40, the reduced dust
content is not surprising.  However, He~2-10 has roughly solar
metallicity \citep{kobulnicky95}, and we would not expect a reduced
dust content.

In general, higher order principle components tend to have
  decreasing signal to noise, but it is evident by eye that components
  3 and 4 are still capturing physically interesting information.  In
  the analysis without the continuum, PCA\#3 indicates the difference
  between CCH and the other lines, highlighting an elongated region,
  also bright in the CCH image, that lies in between the CO-bright E
  cloud and the main body of He 2-10.  That region is captured by
  PCA\#4 in the analysis including continuum.

\subsection{Line Ratios}
\label{Line_Ratios}

While the PCA analysis in Section~\ref{PCA} provides a general census
of the relationship between emission from the molecular lines observed
here, in order to probe how this line emission may depend on local
physical conditions, we determine the various line strengths and
ratios in specific regions.  In particular, we determine the line
intensities in the regions of $0.''8$ radius located around peaks in
the peak intensity maps for each transition (see
Figure~\ref{Mom8_regions} and Table~\ref{line_strengths}).

As discussed in Section~\ref{morphology}, the ratios of different
molecular transitions hold the potential of providing constraints on
the physical conditions due to their differential dependence on
temperature and density.  These line ratios are also dependent on the
relative abundances of each species, which reflects the chemical
pathways (both formation and destruction) in a given physical
environment.  Here we utilize CO(2-1), HCN(1-0), HNC(1-0),
HCO$^+$(1-0), and CCH(1-0), (although we caution the reader that
  many of the HNC(1-0) and CCH(1-0) measurements are relatively weak,
  and have correspondingly large uncertainties.

  Given the critical densities and the energy levels of these
transitions, to first order, we expect the CO(2-1)/HCN(1-0) and
CO(2-1)/HCO$^+$(1-0) ratios to reflect the relative density of the gas.
The ratio of the
HCN(1-0) to HNC(1-0) isomers can also be instructive because these
isomers have similar dipole moments and energy levels; the ratio of
the HCN(1-0) and HNC(1-0) lines can provide insight into chemical
pathways and activation temperatures.  We also expect the
CCH(1-0)/CO(2-1) ratio to track the hard radiation field of
newly-formed massive stars, given that CCH is enhanced by the
photodissociation of CO.  Despite their similar energy levels and
critical densities, the HCN(1-0)/HCO$^+$(1-0) ratio is known to show
strong variations near XDRs (decreasing to $<1$), while it is larger
($>1$) in PDRs if the density is relatively high $> 10^5$
\citep{meijerink07}.

\begin{figure*}
\plottwo{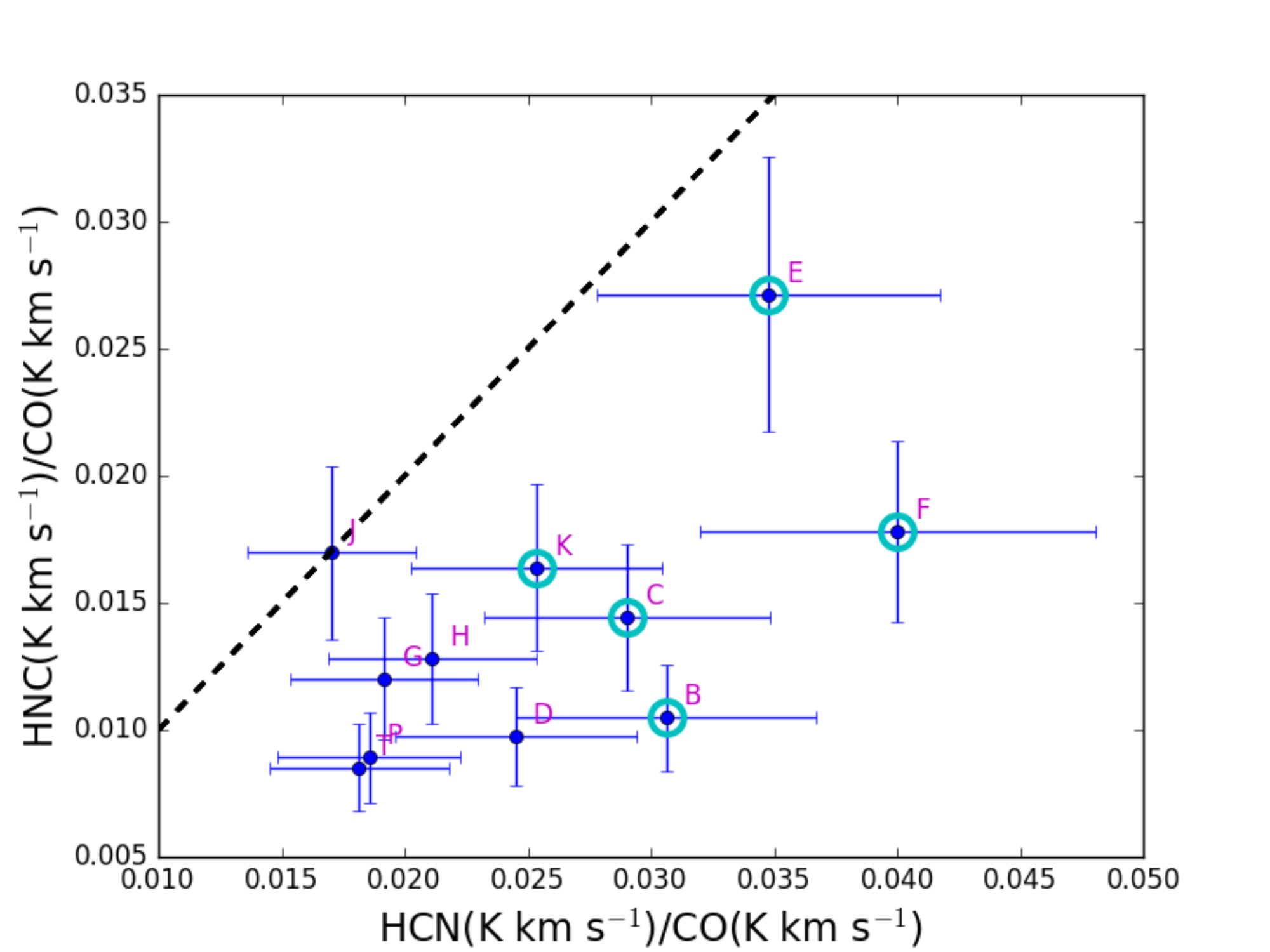}{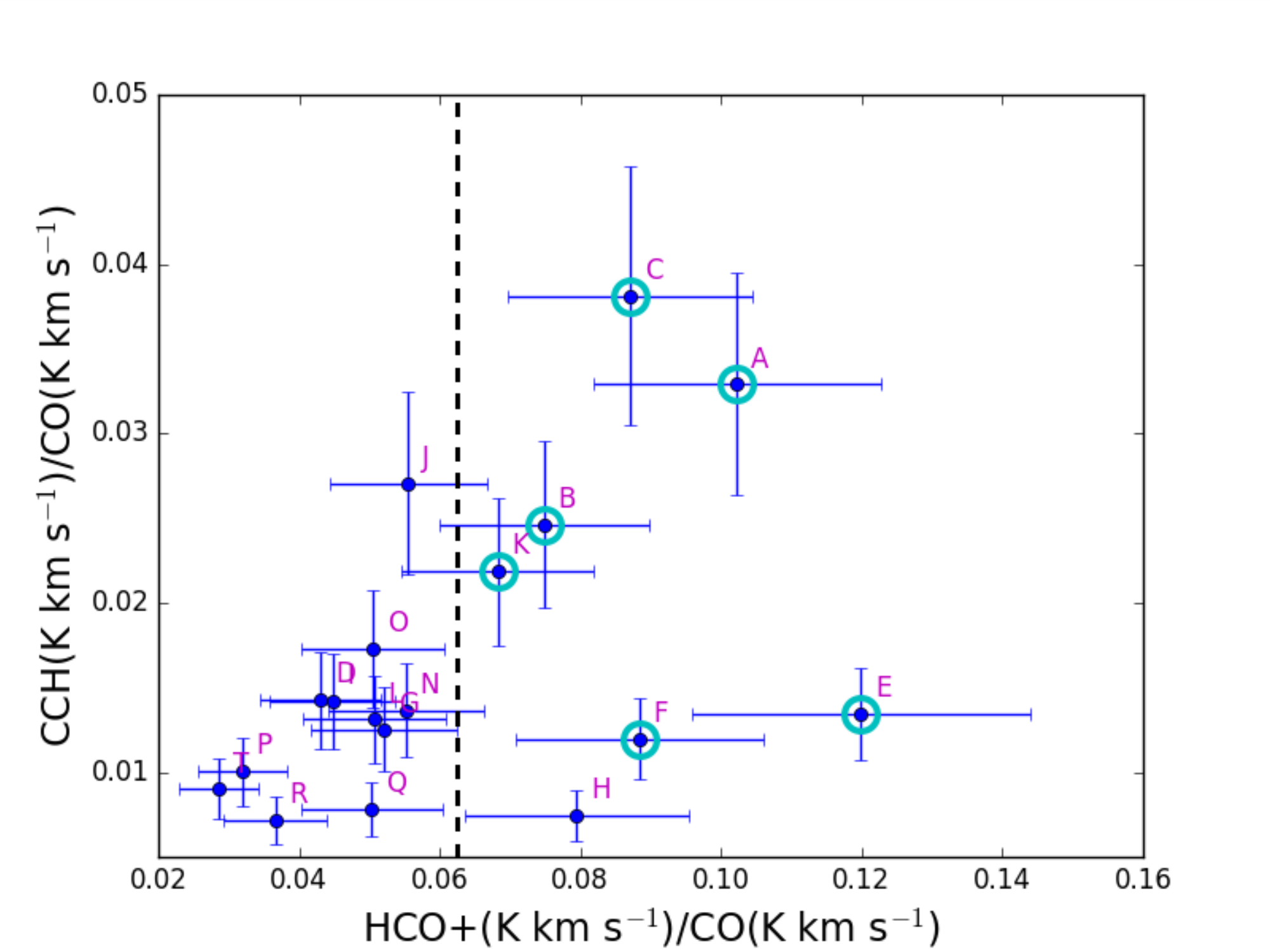}
\caption{Line strengths and ratios for the regions around peaks in the
  integrated intensity map as labeled in Fig.~\ref{Mom8_regions} (see
  Section~\ref{Line_Ratios}). (left) The relationship between
    HCN(1-0) and HNC(1-0) line strengths normalized by the CO(2-1)
    strength. The dashed line corresponds to
  HCN(1-0)$=$HNC(1-0).  (right) The relationship between the
    ratios of HCO$^+$(1-0)/CO(2-1) and CCH(1-0)/CO(2-1). The dashed
  line corresponds to CO(2-1)$= 16\times$HCO$^+$(1-0).  The
cyan-highlighted points indicate regions that overlap with thermal
3.6cm emission (as shown in Fig.~\ref{Mom8_regions}).  Region~C also
includes a low-luminosity AGN \citep{reines11}.  }
\label{plot_line_ratios1}
\end{figure*}

\begin{figure*}
\plottwo{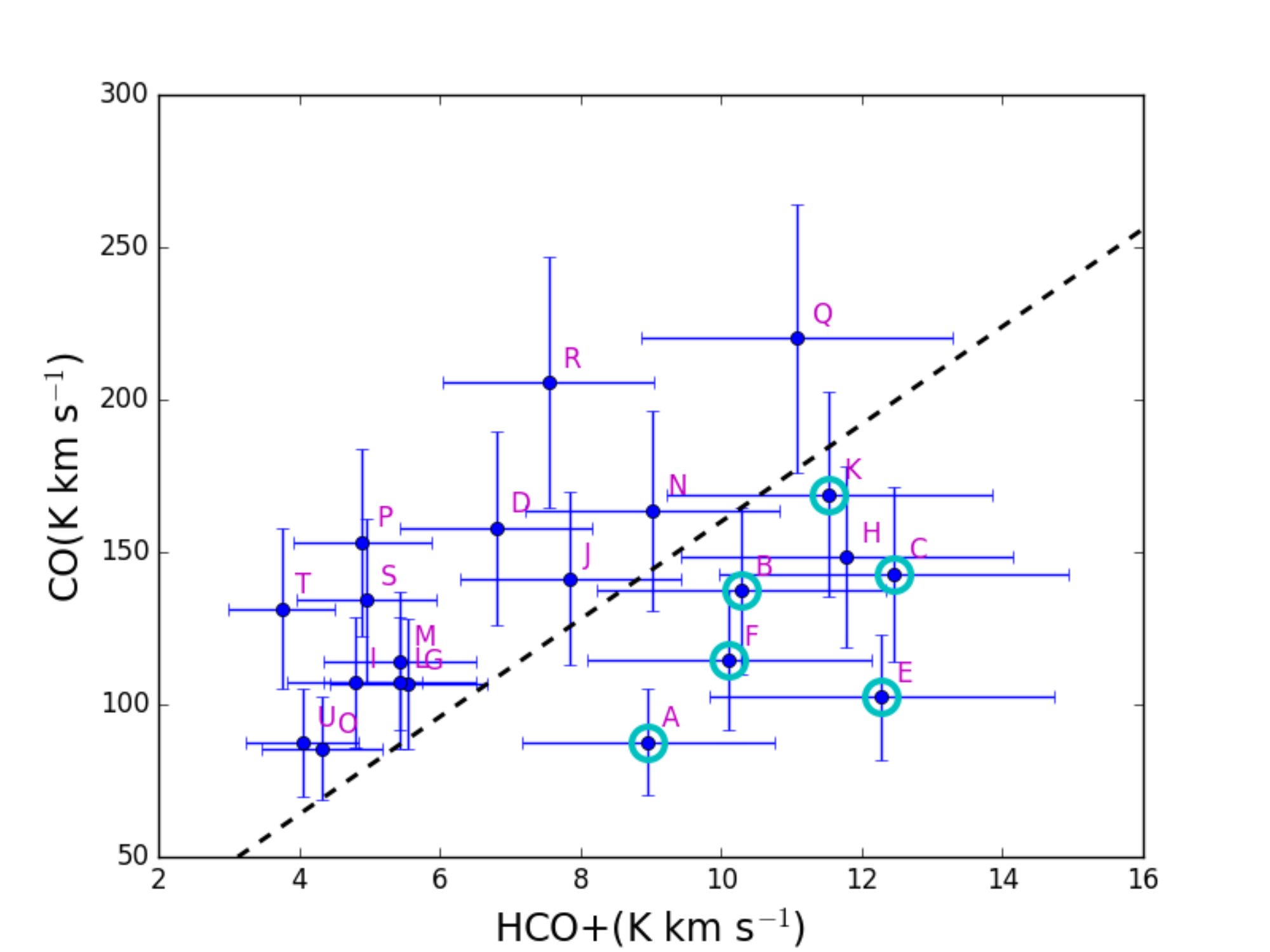}{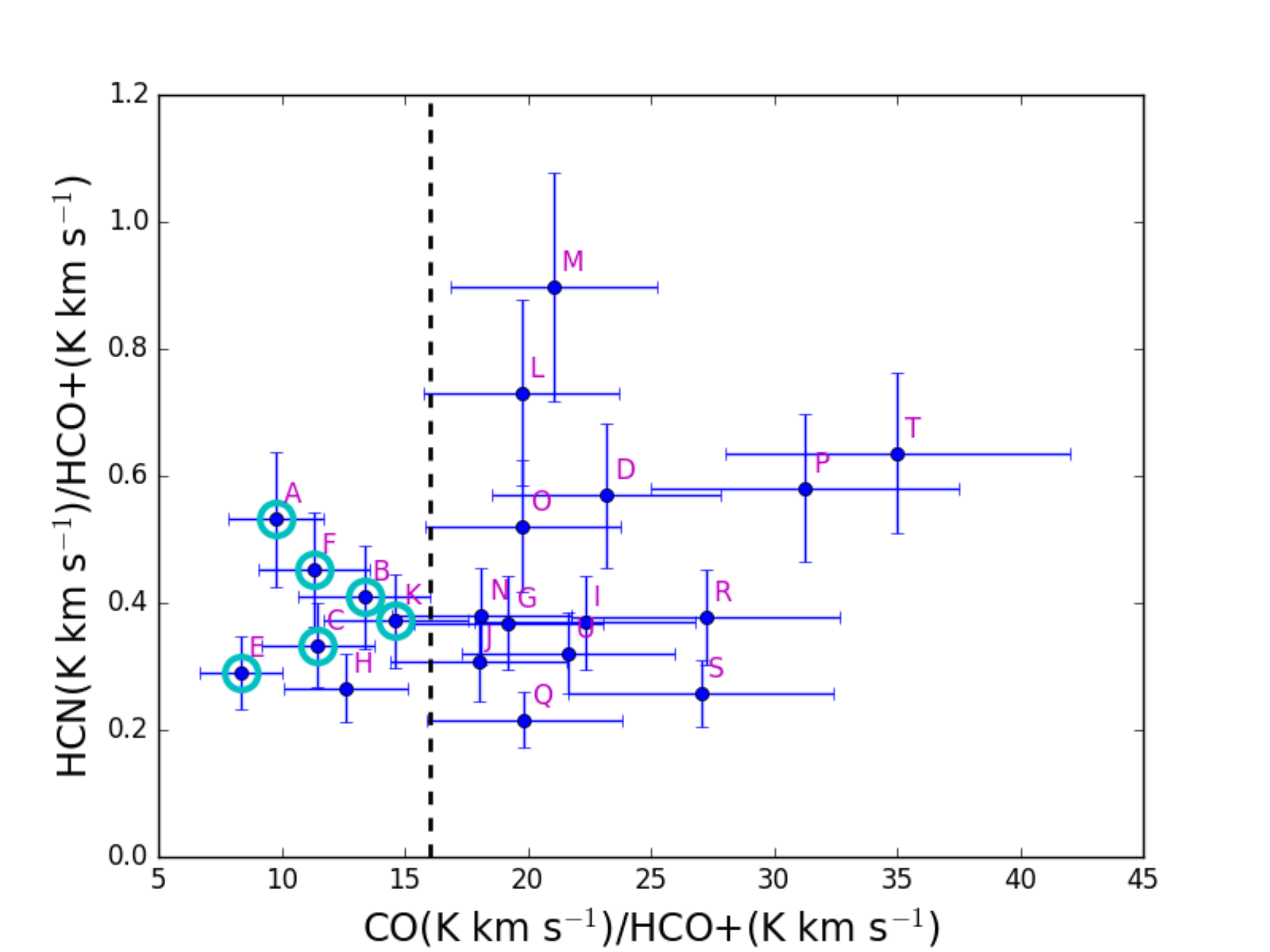}
\caption{ (left) The relationship between the CO(2-1) and HCO$^+$(1-0)
  line strengths.  (right) The HCN(1-0)/HCO$^+$(1-0) ratio as a
  function of the CO(2-1)/HCO$^+$(1-0) line ratio.  The points shown
  are for the regions around peaks in the integrated intensity map as
  labeled in Fig.~\ref{Mom8_regions} (see Section~\ref{Line_Ratios}).
  The cyan-highlighted points indicate regions that overlap with
  thermal 3.6cm emission (as shown in Fig.~\ref{Mom8_regions}).
  Region~C also includes a low-luminosity AGN \citep{reines11}.  The
  dashed line corresponds to CO(2-1)$= 16\times$HCO$^+$(1-0), and cleanly
  separates the sources with and without associated thermal radio
  emission with the exception of Source~H.}
\label{plot_line_ratios2}
\end{figure*}

In Figures~\ref{plot_line_ratios1} and \ref{plot_line_ratios2} we plot
the results of the line ratios (in units of K~km~s$^{-1}$) that
result from the apertures labeled in Figure~\ref{Mom8_regions} that
correspond to peaks in the integrated intensity map in one of the
lines observed here.  We also highlight (in cyan) the apertures that
overlap with the observed thermal radio emission resulting from natal
super star clusters.

A number of trends are apparent in these
Figures~\ref{plot_line_ratios1} and \ref{plot_line_ratios2}, but
perhaps most striking is the extent to which the line strengths
and ratios associated with the natal super star clusters are clearly
distinct.  For example, with the exception of CO, the other line
emission all tends to be stronger near the radio sources.  The radio
sources are particularly well separated in CO/HCO$^+$ as shown in
Figure~\ref{plot_line_ratios2}.  With the exception of region~H, the
line of CO(2-1)$<16\times$HCO$^+$ cleanly separates the line emission
from the radio and non-radio regions.  The HCN/HCO$^+$ ratio is always
$<1$ for all of the regions sampled, and $<0.6$ for the radio sources,
potentially indicating XDRs.   On the other hand the HNC/HCN ratio is
$<1$ for all regions except J, with typical values of $\sim 0.5$ for
the radio sources, indicating that these regions are not particularly
cold but have T$_{\rm kin} \lesssim 50$ ~K \citep{schilke92}.

\subsection{Comparison Between Single Dish and Interferometric Observations}
\label{singledish}

A comparison to single dish observations can be used to determine what
fraction of the total flux we are recovering with these SMA
observations.  Integrating the CO(2-1) emission of these SMA data
within a $1\sigma$ contour around the galaxy results in an integrated
CO(2-1) intensity of $382\pm12$~Jy~km~s$^{-1}$.  \citep{meier01} present
single dish observations of He~2-10 using the CO(3-2) transition with
a total intensity $I_{CO} = 16.6$~K~km~s$^{-1}$ ($=780$~Jy~km~s$^{-1}$
for their beam size of $22''$).  Adopting the brightness temperature
ratio of CO(3-2)/C(2-1)$=0.9$ inferred by \citep{meier01} (equivalent to
$B_{3-2}/B_{2-1} = 1.88$), results {\bf in} a total CO(2-1) intensity of
415~Jy~km~s$^{-1}$.  Thus, these SMA observations appear to recover
$> 90$\% of the CO(2-1) flux from He~2-10.

We can also compare the total HCN(1-0) emission from the ALMA data to
previous observations of \citet{santangelo09} using the IRAM 30~m
  telescope, with a beam size of 28$''$ for the target HCN(1-0) line,
in which they report a single dish integrated flux of
$2.0\pm0.5$~Jy~km~s$^{-1}$.  Using the same $1\sigma$ CO(2-1) contour
as above, we measure a HCN(1-0) integrated flux of
$1.27\pm0.15$~Jy~km~s$^{-1}$, suggesting that these ALMA observations
only recover $\sim 64$\% of the HCN(1-0) emission.  We note that this
fraction of recovered flux is similar to the value found with very
similar ALMA observations by \citet{kepley16} of 70\% for the dwarf
starburst galaxy IIZw40.

It is striking that the ALMA observations of He2-10 appear to recover
a smaller fraction of the HCN(1-0) flux than of the CO(2-1) flux; in
the absence of calibration issues, this result would suggest that the
HCN emission is coming from more extended structures than the CO
emission -- clearly contrary to expectations.  Instead, we offer two
additional potential explanations: 1) The fraction of the CO flux that
appears to  be recovered by ALMA is inflated.  If we instead assume that
the CO(3-2) is fully thermalized, and the CO(3-2)/CO(2-1)$=1$, the
recovered fraction of CO is partially mitigated (bringing the recovered
fraction down to ~80\%).  2) The single-dish flux calibration is too
low.  Given the difficulty in calibrating single-dish observations, it
is not out of the question that this uncertainty could be off by as
much as $\sim 30$\%.  Either or both of these explanations could be
contributing to the apparent discrepancy between the total recovered
flux of CO and HCN.

\subsection{The Gas and Dust Mass}
\label{dust_mass}

\subsubsection{88~GHz Continuum}
As shown in Figure~\ref{4panel}, the 88~GHz continuum emission is
strongly detected throughout the main body of He~2-10, with an
integrated flux density of $F_{88GHz} = 11.4\pm 1.7$~mJy.  The
strongest peaks are co-spatial with known thermal radio sources
(indicating natal super star clusters).

The 88~GHz continuum is likely to be a combination of both dust and
free-free emission.  Based on the radio observations of \citet{jk03},
we can extrapolate the free-free contribution to 88~GHz assuming a
standard spectral index of $\alpha=-0.1$.  We find that $\sim 40$\% of
the 88~GHz emission from the compact regions surrounding the thermal
radio sources is due to dust emission.  If the more diffuse emission
throughout the galaxy has a similar contribution from free-free
emission, that suggests that the 88~GHz emission due to dust is
$4.2$~mJy.  If the more diffuse component has a minimal contribution
from free-free emission, then the total dust emission at 88~GHz due to
dust is $6.5$~mJy.  These values are not inconsistent with the
\citet{santangelo09} observations at 230~GHz, in which the mm
continuum was not detected, and they derived a peak upper limit of
$\sim 5$~mJy.  Extrapolating this upper limit assuming a $\nu^{3.5}$
dust spectral energy distribution suggests the dust contribution at
88~GHz is either small and/or the dust emission is extended.

Adopting a dust contribution to the 88~GHz flux density of
4.2-6.5~mJy, the total dust continuum flux density can be used to
calculate the total gas mass following \citep[][but adopting a
$M_{gas}/M_{dust}=100$]{lh97}:

\begin{equation}
\frac{M_{gas}}{M_\odot} \approx 1.2\times 10^{-16} 
\frac{(\frac{S_\nu}{Jy})(\frac{\lambda}{\mu
    m})^3(\frac{D}{pc})^2}{(\frac{\kappa_{\nu}}{cm^2 g^{-1}})} (e^x-1)
\end{equation}

where, 

\begin{equation}
x = \frac{1.44\times 10^4}{(\frac{\lambda}{\mu m}\frac{T}{K})}
\end{equation}

We assume the emission is optically thin isothermal emission, and a
temperature of 25~K.  Assuming that the dust opacity at 230~Ghz is
$\kappa_{230GHz} = 0.005$~cm$^2$~g$^{-1}$ (per gram of gas), and that
$\kappa(\nu) \propto \nu^\beta$ implies that $\kappa_{88GHz} =
0.0007$~cm$^2$~g$^{-1}$ \citep{andre00}.  
The resulting gas mass is $M_{gas} = 4-6\times 10^{8}$~M$_\odot$.
This value {\bf is} broadly consistent with that inferred by
\citet{kobulnicky95} of $1.6-9.6\times 10^8$~M$_\odot$ based on
CO(1-0) observations.  The {\bf stellar} mass of He~2-10 was inferred
to be $\sim 10^{10}$~M$_\odot$ by \citet{nguyen14} based on the
surface brightness profile, which is $\sim 3\times$ the dynamical mass
of $M_{dyn} = 2.7\times 10^9/sin^2(i)$ found by \citet{kobulnicky95}.
These differences could be attributable to the mass being measured out to different radii; \citet{nguyen14} measure the mass within $\sim 4.3$~kpc, while the measurement of \citet{kobulnicky95} Kobulnicky is within $\sim 2$~kpc.
These results suggest that roughly 5-20\% of the dynamical mass of
He~2-10 is due to the gaseous component, depending on whether one
adopts the results of \citet{kobulnicky95} or \citet{nguyen14}.  This
value might be regarded as a lower limit given the potential for flux
to be resolved out in these interferometric observations, however, as
discussed in Section~\ref{singledish}, these observations appear to
recover $\gtrsim 90$\% of the CO emission compared to single-dish
measurements.  On the other hand, contamination from free-free
emission at 88~GHz would cause the inferred gas mass to be an
overestimate.  Future continuum observations at higher frequency
should help resolve this issue.

\section{Discussion}
 \label{discussion}

\begin{figure*}
\plottwo{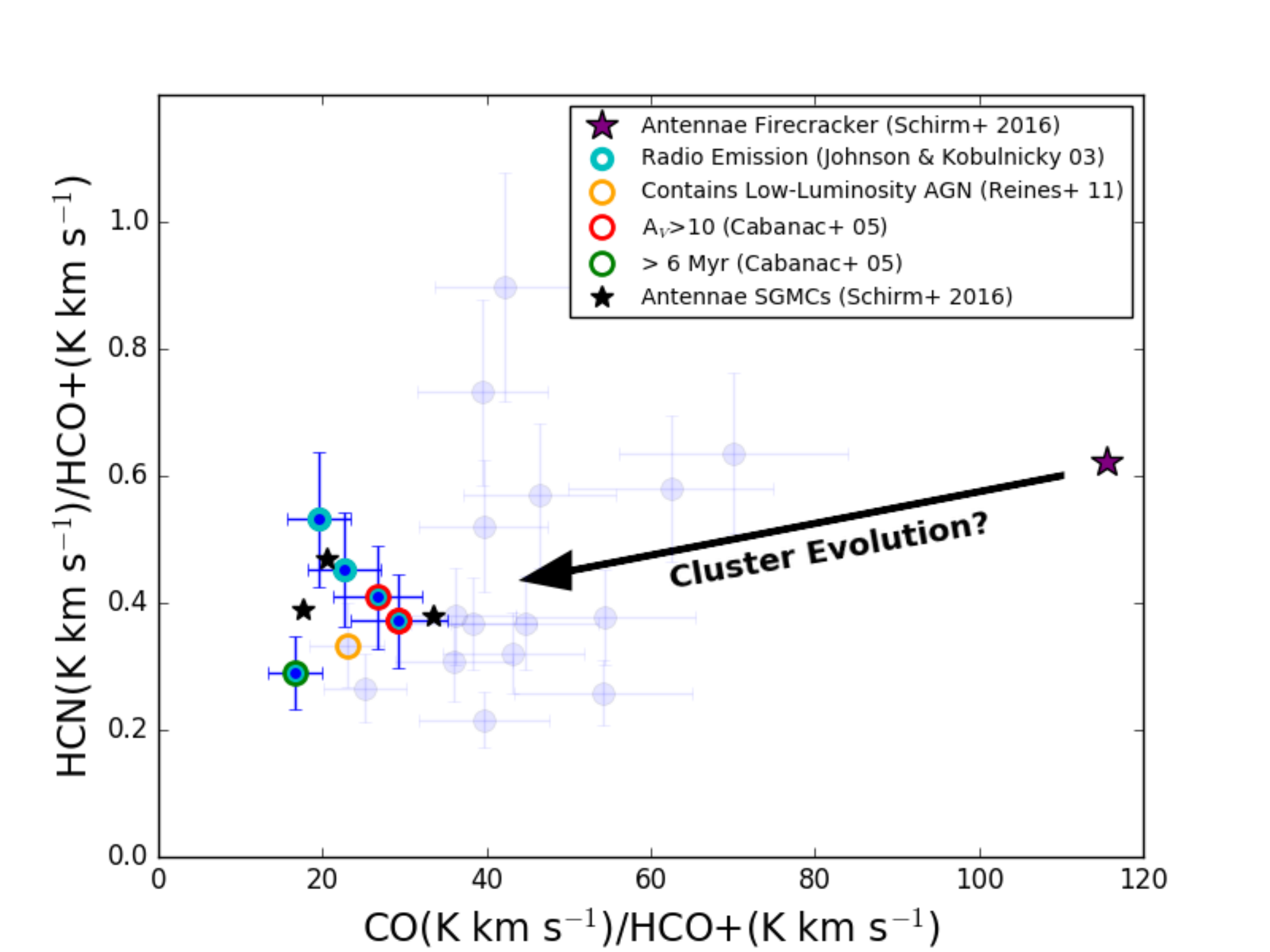}{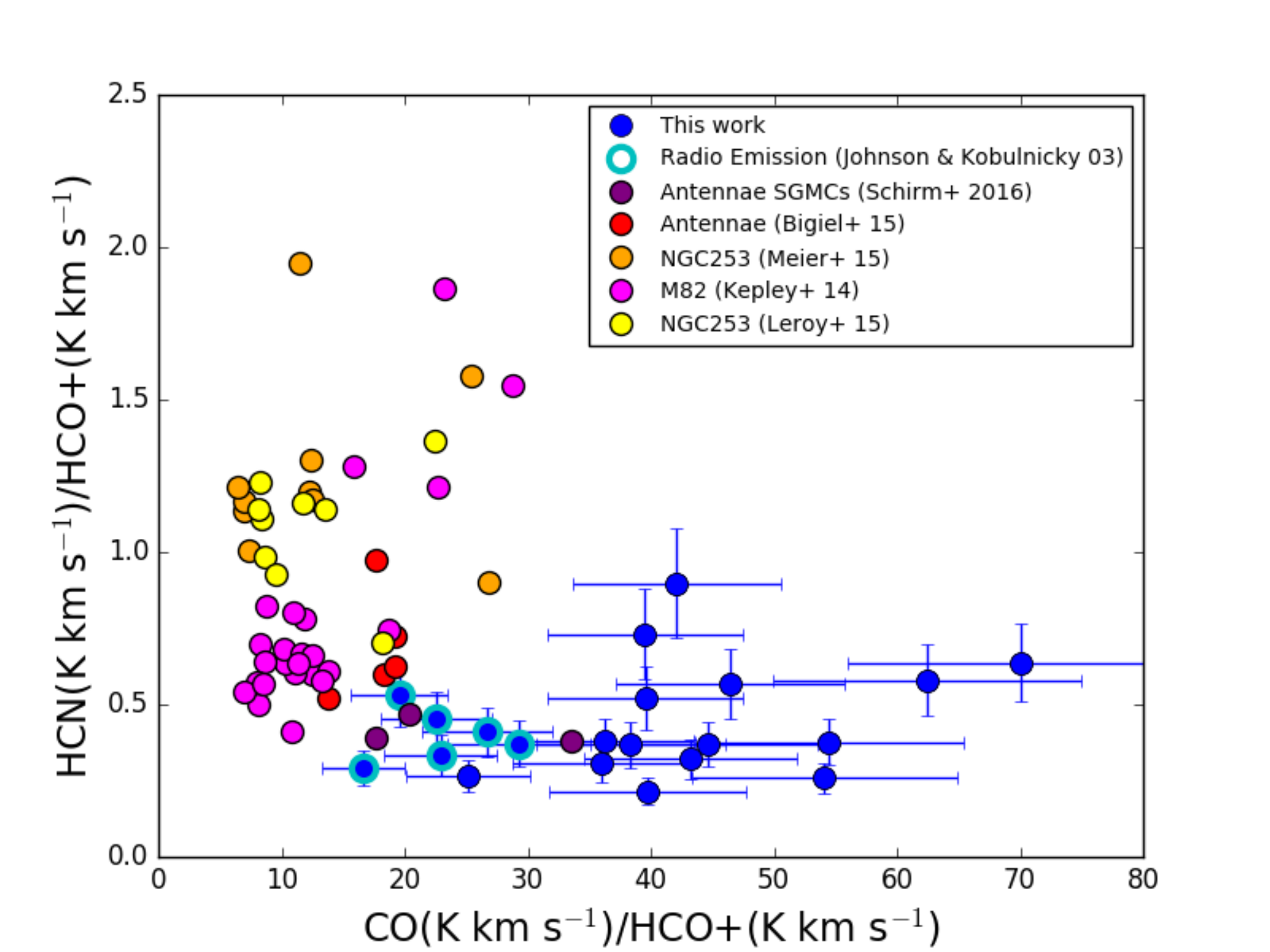}
\caption{Line ratios as shown in Figure~\ref{plot_line_ratios2}.
  (left) Values from regions in He~2-10 and also including the line
  ratios from the ``Firecracker'' region in the Antennae
  \citep{johnson15,schirm16}.  (right) Values from regions in
  comparison galaxies.  For the purposes of comparison, the CO(2-1)
  line strengths have been converted to CO(1-0) assuming
  CO(1-0)/CO(2-1) = 2.  }
\label{plot_comp1}
\end{figure*}

\subsection{Molecular Line Ratios as an Indicator of Cluster Evolution
  \label{evolution}} Given that regions around the thermal radio
sources have clearly distinct line ratios (as shown in
Figures~\ref{plot_line_ratios1} and \ref{plot_line_ratios2}), we
further investigate the evolutionary state of the natal clusters
associated with these regions.  Figure~\ref{plot_comp1} is similar to
Figure~\ref{plot_line_ratios2}b, but in order to facilitate comparison
with other data sets we convert the CO(2-1) line strengths to CO(1-0)
assuming CO(1-0)/CO(2-1)$=2$.  We also now include information about
evolutionary states of specific sources, as well as the location of
the ``Firecracker'' source in the Antennae galaxy system that was
shown by \citet{johnson15} to be a pre-stellar molecular cloud likely
to form a super star cluster in the next $\sim 1$~Myr.  Subsequent
observations by \citet{schirm16} measured the relevant line ratios for
this region, as indicated in Figure~\ref{plot_comp1}, which place it
at a high CO/HCO$^+$ ratio and higher than average HCN/HCO$^+$ ratio
relative to the regions in He~2-10.  We also show the line ratios for
the {\it entire} super giant molecular clouds (SGMCs) in the Antennae,
which are known to already host a large number of star clusters.
Additionally we identify the two natal clusters known to have high
optical extinction \citep[A$_V > 10$][suggesting a very young
evolutionary state]{cabanac05}, and a cluster known to be relatively
old, with an age $>6$~Myr \citep{cabanac05}.

Examining Figure~\ref{plot_comp1} reveals an apparent trend in line
ratios with cluster evolutionary state, with
CO/HCO$^+\propto$~age$^{-1}$. Although this trend could be
  inferred from only the He~2-10 data, it is even more apparent by
  including the ``Firecracker'' region in the Antennae.  The
dependance of the CO/HCO$^+$ ratio might be understood in a few
different ways: 1) The change in this ratio could reflect the density
of the molecular gas.  Given that the critical density of HCO$^+$ is
significantly larger than CO(2-1) \citep[e.g.][]{shirley15},
this scenario could only work if the molecular gas becomes {\it
  denser} as a cluster evolves, which we find unlikely.  2) The HCO$^+$
emission could be enhanced in the PDRs surrounding the young star
clusters \citep[e.g.][]{ginard12}.  This behavior in HCO$^+$
emission has been observed in a number of sources, and therefore we
find this scenario to be plausible.  3) The CO could be subject to
dissociation amid the strong radiation fields and mechanical
luminosity of the young massive stars.  This behavior is predicted by
theory \citep[e.g.][]{kazandjian15}, and therefore we also find
this scenario to be likely.

  The ratio of HCN/HCO$^+$ shows a hint of a trend inversely related to
  age, but this is less clear than the CO/HCO$^+$ trend.  If this
  apparent trend is real, it could be understood as either 1) the
  enhancement of HCO$^+$ in the PDRs driving this ratio down, or 2) the
  density of the gas falling below the critical density of HCN as
  clusters evolve, but staying above the critical density of HCO$^+$.
  \citet{anderson14} found typical values of HCN/HCO$^+$ of $\sim 0.2$
  for parsec-scale molecular clumps not obviously associated with
  current star formation in the 30~Doradus region, which supports the
  apparent trend in this line ratio with evolutionary state presented
  here.  The numerical PDR models of \citet{meijerink07} suggest that the
  ``typical'' HCN/HCO$^+$ ratios for the regions in He~2-10 of $\sim
  0.5$ indicate densities of $\sim 10^4 - 10^5$~cm$^{-3}$.  

\subsection{Comparison to Observations in Other Galaxies}

In order to put these results in context, here we compare the
molecular line ratios presented here to those that have been observed
in other galaxies.  In particular, the relatively nearby starburst
galaxies NGC~253 and M82 provide important benchmarks; not only do
they have high levels of star formation, but their proximity has
enabled studies with relatively small synthesized beams which better
facilitates comparison.  The NCG~253 observations were presented by
\citet{meier15} and \citet{leroy15} and cover the inner $\sim 1$~kpc
of the galaxy, with synthesized beams of $\sim 4''$ and $\sim 2''$,
respectively (corresponding to $\sim 70$~pc and $\sim 35$~pc).  The
M82 observations of \citet{kepley14} have a beam size of $\sim 9''$,
corresponding to $\sim 150$~pc.  The Antennae observations of
\citet{schirm16} have a resolution of $\sim 1.8''$, corresponding to
$\sim 190$~pc, while the observations of \citet{bigiel15} have a
resolution of $\sim 5''$, corresponding to a $\sim 530$~pc.

Comparing these data sets is complicated by their different linear
resolutions, but some trends are still apparent.  For example, with
the exception of the SGMCs in the Antennae, all of the other data
points are in a different region of Figure~\ref{plot_comp1}. In
particular, these regions from other galaxies appear to have
systematically higher HCN/HCO$^+$ ratios, and lower CO/HCO$^+$ ratios.
If the hypothesis presented in Section~\ref{evolution} and
Figure~\ref{plot_comp1} is correct, the location of the points from the
comparison galaxies would suggest that they contain more evolved star
forming regions than the natal clusters sampled in He~2-10.  Given the
location of these regions in the heart of major starbursts, it is not
surprising that the molecular material has been significantly affected
by the intense star formation.  The decreasing CO/HCO$^+$ ratio in
these regions can be understood by the same mechanisms discussed in
Section~\ref{evolution} -- enhancement of HCO$^+$ in PDRs, and
photodissociation of CO.   However, the apparent increase in the
HCN/HCO$^+$ ratio is more puzzling, and we do not at present have an
explanation for this behavior.  The models of \citet{meijerink07}
would suggest that the higher values of HCN/HCO$^+$ correspond to
higher densities, which is contrary to our expectation for the
behavior of the surrounding molecular material as clusters evolve.

\section{Summary}
 \label{summary}

We present ALMA observations of the dwarf starburst galaxy He~2-10
that include the molecular lines of HCO$^+$(1-0), HCN(1-0), HNC(1-0),
and CCH(1-0) as well as 88~GHz continuum.  These observations are
combined with previous CO(2-1) observations obtained with the SMA in
order to characterize the molecular emission of natal super star
clusters.  The main results of this work are the following: 

$\bullet$ Based on continuum measurements at 88~GHz, we infer a total
gas mass of He~2-10 of $M_{gas} = 4-6 \times 10^8$~M$_{\odot}$,
roughly 5-20\% of the dynamical mass. 

$\bullet$ Of the molecular lines observed here, HCO$^+$ most closely
traces the thermal radio emission, suggesting that it is most strongly
correlated with the natal super star clusters. From a principle
component analysis, HCO$^+$ is also the best ``general'' tracer of
molecular emission in He~2-10.

$\bullet$ The line widths and CO luminosities associated with the
regions in He~2-10 suggest that the molecular clouds could have
sizes as small as $\sim 8$~pc, or alternately the line widths are
enhanced, potentially due to a high pressure environment.

$\bullet$ The principle component analysis indicates that the CO
emission and the 88~GHz continuum are {\it anti}-correlated, which
suggests that either the dust and molecular gas components are
not cospatial, or the 88~GHz continuum is dominated by free-free
emission.

$\bullet$ The CO and CCH emission are also anti-correlated, which is
consistent with CCH being enhanced by the photo-dissociated of CO in
the vicinity of the hot young stars in the natal clusters.

$\bullet$ The molecular line ratios of regions containing the natal
star clusters are distinct from the line ratios observed for regions
elsewhere in the galaxy.   In particular, the regions with thermal
radio emission all have CO(2-1)/HCO$^+ < 16$. 

$\bullet$ Based on known properties of the natal
super star clusters in He~2-10, the HCO$^+$/CO ratio appears to be
correlated with the evolutionary stage of the clusters.  This trend
could be due to increasing photo-enhancement of HCO$^+$ as a clusters'
massive stars begin to fully impact their environment, and/or
photodissociation of CO as the high energy radiation from the massive
stars begins to permeate the surrounding regions.

$\bullet$ We suggest that the line ratios in He~2-10, resolved on
typical molecular cloud scales, might help provide insight into more
distant unresolved systems containing embedded star formation.

\acknowledgements 

This research is supported by NSF grant 1413231 (PI: K. Johnson).
This paper makes use of the following ALMA data:
ADS/JAO.ALMA\#2011.0.00348.S ALMA is a partnership of ESO
(representing its member states), NSF (USA) and NINS (Japan), together
with NRC (Canada), NSC and ASIAA (Taiwan), and KASI (Republic of
Korea), in cooperation with the Republic of Chile. The Joint ALMA
Observatory is operated by ESO, AUI/NRAO and NAOJ.  The National Radio
Astronomy Observatory is a facility of the National Science Foundation
operated under cooperative agreement by Associated Universities, Inc..
This work was partly supported by the Italian Ministero
dell'Istruzione, Universit\`a e Ricerca through the grant Progetti
Premiali 2012 -- iALMA (CUP C52I13000140001). A.E.R. is grateful for
support from NASA through Hubble Fellowship grant HST-HF2-51347.001-A
awarded by the Space Telescope Science Institute, which is operated by
the Association of Universities for Research in Astronomy, Inc., for
NASA, under contract NAS 5-26555.

\end{document}